\documentclass[10pt,journal,compsoc, review]{IEEEtran}

\usepackage{amsthm}
\usepackage{amsfonts,amsmath}
\usepackage{algorithm}
\usepackage[noend]{algpseudocode}
\usepackage{bm}
\usepackage{booktabs}
\usepackage{soul}
\usepackage{multirow}
\usepackage{subfig}
\usepackage[flushleft]{threeparttable}
\usepackage{adjustbox}
\usepackage{fancyhdr}
\usepackage[dvipsnames]{xcolor}
\usepackage[flushleft]{threeparttable}
\usepackage{tikz}
\usepackage{outlines}
\usepackage{balance}
\usepackage{tcolorbox}
\usepackage{wrapfig}
\usepackage{blindtext}
\usepackage{xcolor}
\usepackage{balance}
\usepackage{algorithm}
\usepackage{color}
\usepackage{listings}
\usepackage{color}
\usepackage{MnSymbol,wasysym}
\usepackage{marvosym}
\usepackage[switch]{lineno}

\definecolor{codegreen}{rgb}{0,0.6,0}
\definecolor{codegray}{rgb}{0.5,0.5,0.5}
\definecolor{codepurple}{rgb}{0.58,0,0.82}
\definecolor{backcolour}{rgb}{0.95,0.95,0.92}

\lstdefinestyle{mystyle}{
  backgroundcolor=\color{backcolour}, commentstyle=\color{codegreen},
  keywordstyle=\color{magenta},
  numberstyle=\tiny\color{codegray},
  stringstyle=\color{codepurple},
  basicstyle=\ttfamily\footnotesize,
  breakatwhitespace=false,         
  breaklines=true,                 
  captionpos=b,                    
  keepspaces=true,                 
  numbers=left,                    
  numbersep=5pt,                  
  showspaces=false,                
  showstringspaces=false,
  showtabs=false,                  
  tabsize=2
}
\newcommand{\colorline}[1]{
\hspace{-0.03\linewidth}\colorbox{gray!30}{\makebox[0.99\linewidth][l]{Hu}}
 }

\newtheorem{theorem}{Theorem}
\newtheorem*{proof*}{Proof}

\newtheorem{definition}{Definition}

\ifCLASSOPTIONcompsoc
  \usepackage[nocompress]{cite}
\else
  \usepackage{cite}
\fi

\usepackage{algorithm}
\usepackage{algpseudocode} 

\usepackage{twoopt}
\usepackage{tikz}
\usetikzlibrary{fit}

\newcommand\tikzmark[1]{%
  \tikz[remember picture,overlay]\node[inner xsep=0pt] (#1) {};}
\newcommandtwoopt\Textbox[5][2.5cm][2cm]{%
\begin{tikzpicture}[remember picture,overlay]
  \coordinate (aux) at ([xshift=#1]#4);
  \node[inner ysep=5pt,yshift=0ex,draw=black,
    fit=(#3) (aux),baseline] 
    (box) {};
  \node[text width=#2,anchor=north east,
    font=\sffamily\footnotesize,align=right] 
    at (box.north east) {#5};
\end{tikzpicture}%
}
\DeclareCaptionSubType*{algorithm}

\DeclareCaptionLabelFormat{alglabel}{Alg.~#2}

\begin{document}
\bstctlcite{IEEEexample:BSTcontrol}

\title{GraphAGILE: An FPGA-based Overlay Accelerator for Low-latency GNN Inference}

\author{Bingyi Zhang,
        Hanqing Zeng,
        Viktor Prasanna,~\IEEEmembership{Fellow,~IEEE}
        
\IEEEcompsocitemizethanks{\IEEEcompsocthanksitem B. Zhang, H. Zeng and V. K. Prasanna are with the Department of Electrical and
Computer Engineering, University of Southern California, Los Angeles,
CA 90089. 
E-mail: \{bingyizh, zengh, prasanna\}@usc.edu}
}

\IEEEtitleabstractindextext{%
\begin{abstract}
This paper presents GraphAGILE, a domain-specific FPGA-based overlay accelerator for graph neural network (GNN) inference. GraphAGILE consists of (1) \emph{a novel unified architecture design} with an \emph{instruction set}, 
and (2) \emph{a compiler} built upon the instruction set that can quickly generate optimized code. 
Due to the proposed instruction set architecture (ISA) and the compiler,  GraphAGILE does not require any FPGA reconfiguration when performing inference on various GNN models and input graphs.
For the architecture design, we propose a novel hardware module named Adaptive Computation Kernel (ACK), that can execute various computation kernels of GNNs,
including general matrix multiplication (GEMM), sparse-dense matrix multiplication (SpDMM), and sampled dense-dense matrix multiplication (SDDMM).
The compiler takes the specifications of a GNN model and the graph meta data (e.g., the number of vertices and edges) as input,  and generates a sequence of instructions for inference execution.   
We develop the following compiler optimizations to reduce inference latency: (1) computation order optimization that automatically reorders the computation graph to reduce the total computation complexity, (2) layer fusion that merges adjacent layers to reduce data communication volume, (3) data partitioning with a partition-centric execution scheme that partitions the input graph to fit the available on-chip memory of FPGA,    (4) kernel mapping that automatically selects execution mode for ACK, and performs task scheduling to overlap computation with data communication and achieves dynamic load balance. We implement GraphAGILE on a state-of-the-art FPGA platform, Xilinx Alveo U250. GraphAGILE can execute widely used GNN models, including GCN, GAT, GIN, GraphSAGE, SGC and other GNN models supported by GraphGym. Experimental results show that GraphAGILE achieves up to $47.1\times$ ($3.9\times$) reduction in end-to-end latency, including the latency of compilation and hardware execution, compared with the state-of-the-art implementations on CPU (GPU), and achieves up to  $2.9\times$ reduction in hardware execution latency compared with the state-of-the-art FPGA accelerators.

\end{abstract}

}

\maketitle

\IEEEdisplaynontitleabstractindextext

\IEEEpeerreviewmaketitle

\section{Introduction}
\label{sec:introduction}

\IEEEPARstart{G}{raph} neural networks (GNNs) have achieved unprecedented success in graph-based machine learning. Compared with traditional algorithms, GNNs achieve superior performance for a wide variety of applications \cite{hu2020open}, such as recommendation systems, social media \cite{lerer2019pytorch}, etc. \emph{Low-latency GNN inference} is needed in many real-world applications. 
Examples include real-time traffic prediction \cite{jiang2021graph}, and GNN-based scientific simulation \cite{pfaff2020learning}.  

Accelerating GNN inference is challenging because GNN inference \cite{yan2020hygcn,zhang2021boostgcn,zeng2020graphact} requires both sparse and dense computation kernels. While the sparse computation kernels result in poor data reuse and irregular memory access patterns, the dense computation kernels can be executed with regular memory access patterns. General purpose processors (e.g., CPU, 
GPGPU) are inefficient for GNN inference due to (1) complex cache hierarchy that results in ineffective on-chip memory utilization due to the poor spatial and temporal locality,  (2) the general microarchitecture designs are inefficient for various computation kernels in GNNs (i.e., GEMM, SpDMM, {and} SDDMM).
For GPUs, the state-of-the-art GNN frameworks (e.g., Pytorch Geometric (PyG) \cite{fey2019fast}, Deep Graph Library (DGL) \cite{wang2019deep}) have large inference latency due to (1) large GPU kernel launch time, and (2) sub-optimal execution paradigm for sparse computation leading to large memory traffic. For example, due to the large GPU global memory footprint for storing the intermediate results, programs written with PyG spend 55\%-99\% \cite{yan2020hygcn} time executing the sparse computations of GNN inference.

Many GNN accelerators \cite{zhang2020hardware, yan2020hygcn, geng2020awb, zhang2021boostgcn, liang2020deepburning, zeng2020graphact, lin2021hp, geng2021gcn, auten2020hardware} have been proposed to overcome the inefficiency of  CPUs and GPUs. Previous works either directly design accelerators for specific GNN models \cite{zhang2020hardware, geng2020awb} or develop design automation frameworks \cite{zhang2021boostgcn, lin2021hp, liang2020deepburning} to generate FPGA accelerators for a specific GNN model and an input graph. However, the design automation frameworks need to regenerate optimized hardware design if the structure of the GNN model or the topology of the input graph changes. The hardware regeneration requires meta compilation, hardware synthesis, place\&route, and FPGA reconfiguration, which incur significant overhead and are not suitable for cloud-based FPGA accelerators. A typical end user may explore a variety of GNN models and perform inference on various input graphs. Moreover, in a cloud-based system, multiple users share the same FPGA. Different users may run different GNN models with different input graphs.  
Therefore, the time-consuming process of regenerating an optimized accelerator makes the design automation frameworks unattractive in the above scenarios.


In this paper, we propose an FPGA-based overlay accelerator, GraphAGILE. An FPGA overlay \cite{yu2019opu,abdelfattah2018dla} consists of an instruction set architecture (ISA) and a compiler, providing software-like programmability and targeting a specific application domain. The ISA of GraphAGILE unifies the execution of both the sparse and dense computation kernels of GNNs without hardware reconfiguration.
To program the ISA, the compiler takes as inputs the specification of the GNN model and the graph meta data, and generates a sequence of instructions to execute on the ISA of the overlay architecture. To reduce the inference latency, we propose several optimizations for the compiler to efficiently utilize the ISA. To the best of our knowledge, GraphAGILE is the first FPGA overlay accelerator for GNNs. We summarize our main contributions as follows:

\begin{itemize}
    \item We propose an instruction set architecture to accelerate GNN inference. It supports a broad range of GNN models by efficiently executing various computation kernels in GNNs, including GEMM, SpDMM, {and} SDDMM.  
    \item We develop a compiler that generates an instruction sequence based on an input graph and GNN model. Compiler optimizations include:
    \begin{itemize}
        \item computation order optimization that automatically reorders the computation graph to reduce the total computation complexity.
        \item layer fusion that merges adjacent layers to communicate the inter-layer results through on-chip memory, which reduces the total volume of external memory communication. 
        \item graph partitioning that optimizes the intra-layer and inter-layer data communication under a given on-chip memory constraint.
        \item kernel mapping and task scheduling that hide data communication latency and achieve dynamic load balance.
    \end{itemize}
    \item We deploy GraphAGILE on Xilinx Alveo U250, a state-of-the-art cloud-based  FPGA platform. 
    \begin{itemize}
        \item We demonstrate that GraphAGILE can execute widely used GNN models, including GCN, GAT, GIN, GraphSAGE, SGC, and other GNN models supported by GraphGym \cite{you2020design}.
        \item GraphAGILE achieves up to $47.1\times$ ($3.9\times$) speedup in end-to-end latency (see section \ref{sec:Evaluation-Results}) compared with  the state-of-the-art  implementations on CPU (GPU), and up to $2.9\times$ speedup in hardware execution latency compared with the state-of-the-art FPGA accelerators.
    \end{itemize}
\end{itemize}

{ 
The rest of the paper is organized as follows: Section \ref{Sec:background} introduces the background of graph neural networks; Section \ref{Sec:related-work} covers the related work; Section \ref{Sec:overview} presents an overview of GraphAGILE; Section \ref{Sec:Microarchitecture} describes the microarchitecture design of GraphAGILE; Section \ref{sec:compiler} covers the details of the compiler design; Section \ref{Sec:Experimental-Evaluation} describes the implementation details and Section \ref{sec:Evaluation-Results} includes the evaluation results. 
}

\section{Background}
\label{Sec:background}
{This section introduces the background of graph neural networks  and briefly describes two well-known graph neural network models.}

\subsection{Graph Neural Networks}
\label{subsec:gnn}

\begin{table}[ht]
\centering
\caption{Notations}
\begin{adjustbox}{max width=0.48\textwidth}
\begin{tabular}{cc|cc}
\toprule
 \textbf{{Notation}} & \textbf{{Description}}  & \textbf{{Notation}}  & \textbf{{Description}} \\
 \midrule
\midrule
{$  \mathcal{G}(\mathcal{V},\mathcal{E})$ }& {input graph}  &  $ v_{i}$ & {$i^{th}$ vertex} \\ \midrule
$ \mathcal{V}$ &  {set of vertices} &  $ e(i,j)$ & {edge from $ v_{i}$ to $  v_{j}$} \\ \midrule
$ \mathcal{E}$& {set of edges} &  $ L$&{number of GNN layers} \\ \midrule
$ \bm{h}_{i}^{l}$& feature vector of $ v_{i}$
at layer $l$    &  $ \bm{m}_{i}^{l}$ & aggregated message by vertex $v_{i}$ \\ 

 \bottomrule
\end{tabular}
\end{adjustbox}
\label{tab:notations}
\end{table}

Table \ref{tab:notations} defines the notations in GNN layer operations. 
Graph neural networks (GNNs) \cite{kipf2016semi, hamilton2017inductive} are proposed for representation learning on a graph $  \mathcal{G}(\mathcal{V},\mathcal{E})$. Each edge in $\mathcal{G}$ is associated with a weight.  A GNN model consists of a stack of GNN layers. Each GNN layer performs message passing on $ \mathcal{G}$ where each vertex aggregates information from its neighbors. Thus, a multi-layer GNN model recursively performs such message passing on multi-hop neighbors. According to \cite{fey2019fast, wang2019deep},  a GNN layer can be abstracted as:
\begin{equation}
    \text{Edge-wise}: \bm{m}_{e}^{l} = \phi(\bm{h}_{u}^{l-1}, \bm{h}_{v}^{l-1}, w^{l-1}_{e}),  \forall  e(u,v) \in \mathcal{E} \label{eq:edge-ise}
\end{equation} 
\begin{equation}
    \text{Node-wise}: \bm{h}_{v}^{l} = \psi(\bm{h}_{v}^{l-1}, \rho(\{\bm{m}^{l}_{e}: e(u,v)\in \mathcal{E}\}))  \label{eq:node-ise}
\end{equation}
where $\phi()$ is the \emph{message function}. Each edge uses $\phi()$  to generate a message by combining the edge weight $w^{l-1}_{e}$ with the features of its incident vertices.  $\psi()$ is the \emph{update function}. Each vertex uses $\psi()$ to update its features by aggregating the incoming messages using the \emph{reduction function} $\rho()$. In GNNs, the message/update functions are parameterized by neural network modules \cite{hamilton2017inductive}, such as Multi-layer Perception.
Some well-known GNN models include:

 \textbf{GCN} \cite{kipf2016semi}: each layer is defined as
\begin{equation}
    \begin{split}
        \bm{m}_{i}^{l} & = \text{Sum}\left( \left\{ \alpha_{ji} \cdot \bm{h}_{j}^{l-1}:j\in \mathcal{N}(i)\cup  \{i\}\right\}\right)\\
        \bm{h}_{i}^{l} & = \text{ReLU} \left(\bm{m}_{i}^{l}\bm{W}^{l} \right)
    \end{split}
    \label{label:gcn}
\end{equation}
where $l$ denotes the $l^{\text{th}}$ layer, $\alpha_{ji}=\frac{1}{ \sqrt{D(j)\cdot D(i)}}$ ($D(j)$ is the degree of $v_{j}$), $\bm{W}^{l}$ denotes the weight matrix of layer $l$, and $\mathcal{N}(i)$ denotes the set of neighbors of $v_{i}$.

\textbf{GAT} \cite{velivckovic2017graph}: this model has similar layer definition as GCN. In addition, GAT applies the attention mechanism to calculate edge weight  $\alpha_{ij}$ dynamically:
\begin{equation}
    \alpha_{ij} = \frac{\text{exp}\left(\text{LReLU}\left( \langle\bm{a}_{\text{att}}, [\bm{W}_{\text{att}}\bm{h}_{i}||\bm{W}_{\text{att}}\bm{h}_{j}] \rangle\right)\right)}{\sum_{k\in \mathcal{N}(i)} \text{exp}\left(\text{LReLU}\left( \langle\bm{a}_{\text{att}}, [\bm{W}_{\text{att}}\bm{h}_{i}||\bm{W}_{\text{att}}\bm{h}_{k}] \rangle\right)\right)} \label{eq:gat-attention}
\end{equation}
where $\bm{a}_{\text{att}}$ is an attention vector, $\bm{W}_{\text{att}}$ is an attention matrix, and $ \langle, \rangle$ is the vector inner product operator.







In addition, many other GNN models (e.g., GIN \cite{xu2018powerful}) have been proposed following the recursive message-passing paradigm. Recently, \cite{you2020design} proposes the GraphGym library \cite{you2020design} and defines the general design space of a GNN. The design space includes intra-layer design and inter-layer design, where the intra-layer design follows the message-passing paradigm defined in Equations \ref{eq:edge-ise}, \ref{eq:node-ise}, the inter-layer design adds the residual connections across the GNN layers.

\section{Related Work}
\label{Sec:related-work}


There have been many FPGA overlay accelerators proposed for CNNs targeting image-related tasks, such as AMD Xilinx DPU \cite{xilinxdpu}, Intel DLA \cite{abdelfattah2018dla}, Nvidia NVDLA \cite{Nividanvdla}, TVM VTA \cite{moreau2019hardware}, and OPU \cite{yu2019opu}. These CNN overlay accelerators have similar components: (1) a general architecture design with an instruction set, and (2) a compiler that generates the instruction sequence for the target CNN model.
Compared with CNN overlay accelerators, the design of GNN overlay accelerators is more challenging: (1) The major computation kernel in CNNs is convolution, which can be efficiently supported by a single hardware design, such as a 2-D systolic array. In contrast, GNNs have heterogeneous computation kernels (e.g., GEMM, SpDMM, {and} SDDMM), making it more challenging to design a unified hardware architecture. (2) For a CNN overlay accelerator, the compiler processes images of regular shapes. For tasks such as image classification, CNN models accept input images of a fixed size, and the compiler only needs to generate a single instruction sequence for these models. In contrast, the input graphs to GNNs are independent of GNN models, and the real-world graphs have various sizes and connectivity. The graphs with the same number of vertices and edges may have highly different structures. Thus, the software compiler of the GNN overlay accelerator needs to process graphs of various sizes and connectivity.  
Complex data-dependent optimizations (e.g., complex graph partitioning) in compiler result in large overhead at compilation time, which may degrade the end-to-end latency  (see Section \ref{sec:Evaluation-Results}).      

\vspace{0.2cm}
 { In our prior work \cite{zhang2020hardware}, we proposed a hybrid hardware architecture to accelerate graph convolutional network (GCN) \cite{kipf2016semi}. In \cite{zhang2021boostgcn}, we proposed a partition-centric execution scheme to accelerate GCN by improving the memory performance. In \cite{zhang2022low}, we proposed a unified hardware architecture that supports GEMM and SpDMM in GNNs. Based on our previous works \cite{zhang2020hardware, zhang2021boostgcn, zhang2022low}, we  propose (1) a unified hardware architecture to support three key computation kernels in GNNs, including GEMM, SpDMM, and SDDMM, (2) a general instruction set that enables software-like programmability and supports a broad range of GNNs, and (3) a compiler with latency reduction optimizations (including the computation order optimization in \cite{zhang2020hardware}, and the partition-centric execution scheme in \cite{zhang2021boostgcn}) to automatically compile  the GNN model into an instruction sequence for hardware execution.}
\section{Overview}
\label{Sec:overview}
{ In this section, we introduce the GraphAGILE workflow (Section 4.1) and provide an overview of the hardware architecture (Section 4.2).}

\subsection{Overview of GraphAGILE}
\label{subsec:overview-of-GraphAGILE}

\begin{figure}[h]
     \centering
     \includegraphics[width=8.5cm]{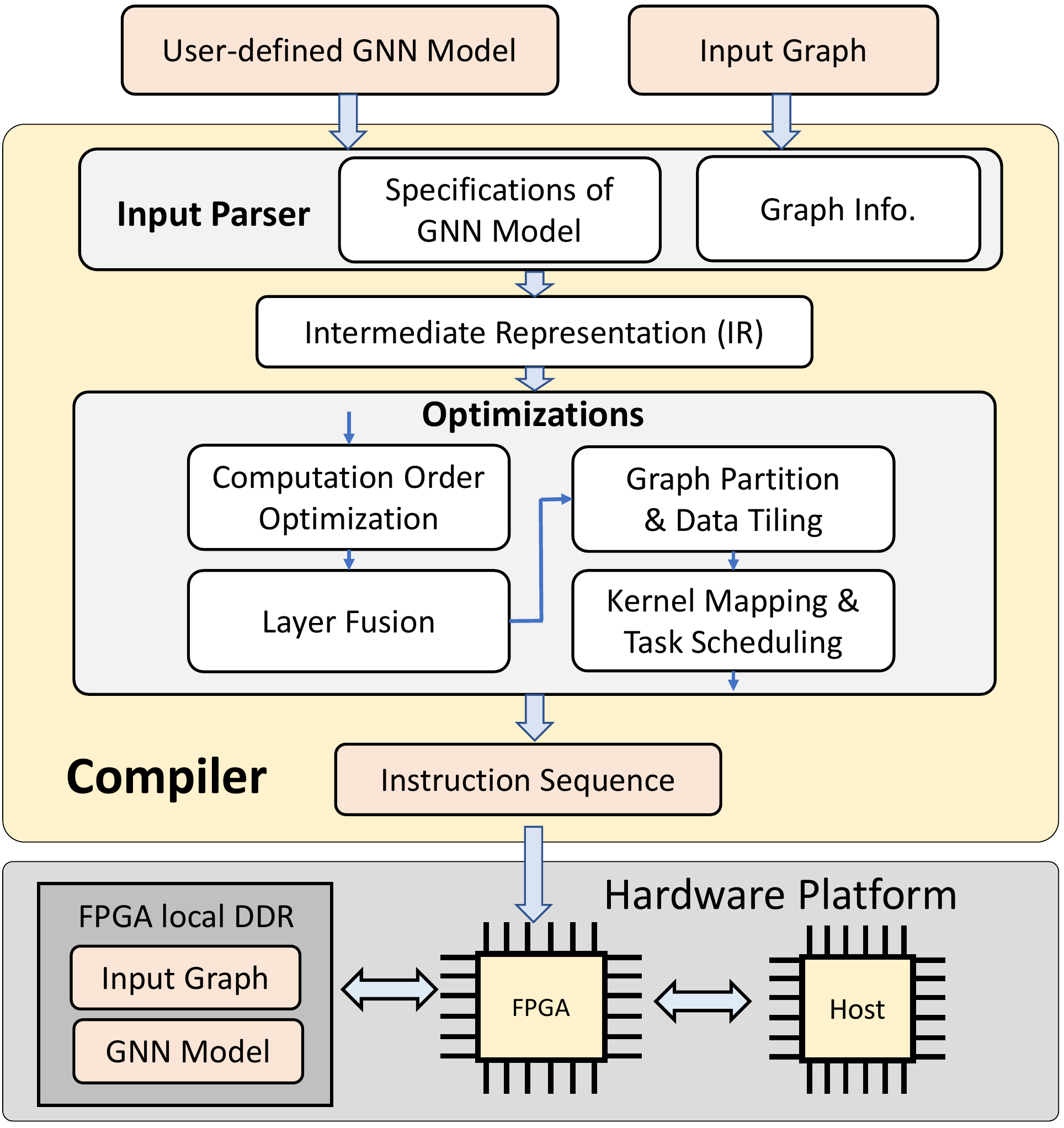}
     \caption{Overview of GraphAGILE}
     \label{fig:overview}
\end{figure}

\vspace{0.1cm}
\noindent \textbf{Target application domain}: This work targets the inference process of various GNN-based applications, such as recommendation system \cite{hamilton2017inductive}, social media, citation networks \cite{kipf2016semi}, etc. In the target applications, the graphs can be very large. For example, a graph in recommendation systems may contain billions of vertices and edges. GraphAGILE supports a broad range of GNN models, including (1) widely used GNN models (GCN \cite{kipf2016semi}, GraphSAGE \cite{hamilton2017inductive}, GAT \cite{velivckovic2017graph}, GIN \cite{xu2018powerful}, SGC \cite{wu2019simplifying}), (2) GNN models in the design space of GraphGym \cite{you2020design}. In addition, GraphAGILE has the potential to be applied to other GNN models. An \emph{instance} to GraphAGILE is specified by (1) the specifications of a GNN model, (2) the specifications of an input graph.

\vspace{0.1cm}
\noindent \textbf{Hardware platform}: The hardware platform consists of an FPGA device,  FPGA local DDR memory, and a host processor. 
The proposed hardware accelerator is deployed on the FPGA device. FPGA local DDR memory stores the input graph, the GNN model, and binary files generated by the compiler.  The compiler is executed on the host processor.

\vspace{0.1cm}
\noindent \textbf{Compiler}: Users define the GNN using Pytorch Geometric (PyG) library. The inputs to the compiler are (1) the computation graph of the GNN model generated by PyG, and (2) the input graph. The Input Parser (Figure \ref{fig:overview}) extracts the specifications of the GNN model and the information of the input graph to generate the \emph{Intermediate Representation (IR)}. After obtaining IR, the compiler performs the four optimization steps on the GNN computation graph as shown in Figure \ref{fig:overview}. Then, the compiler generates a sequence of instructions to execute on the hardware accelerator.

\subsection{Architecture Overview}
\label{subsec:Architecture Overview}

\begin{figure}[h]
     \centering
     \includegraphics[width=8.5cm]{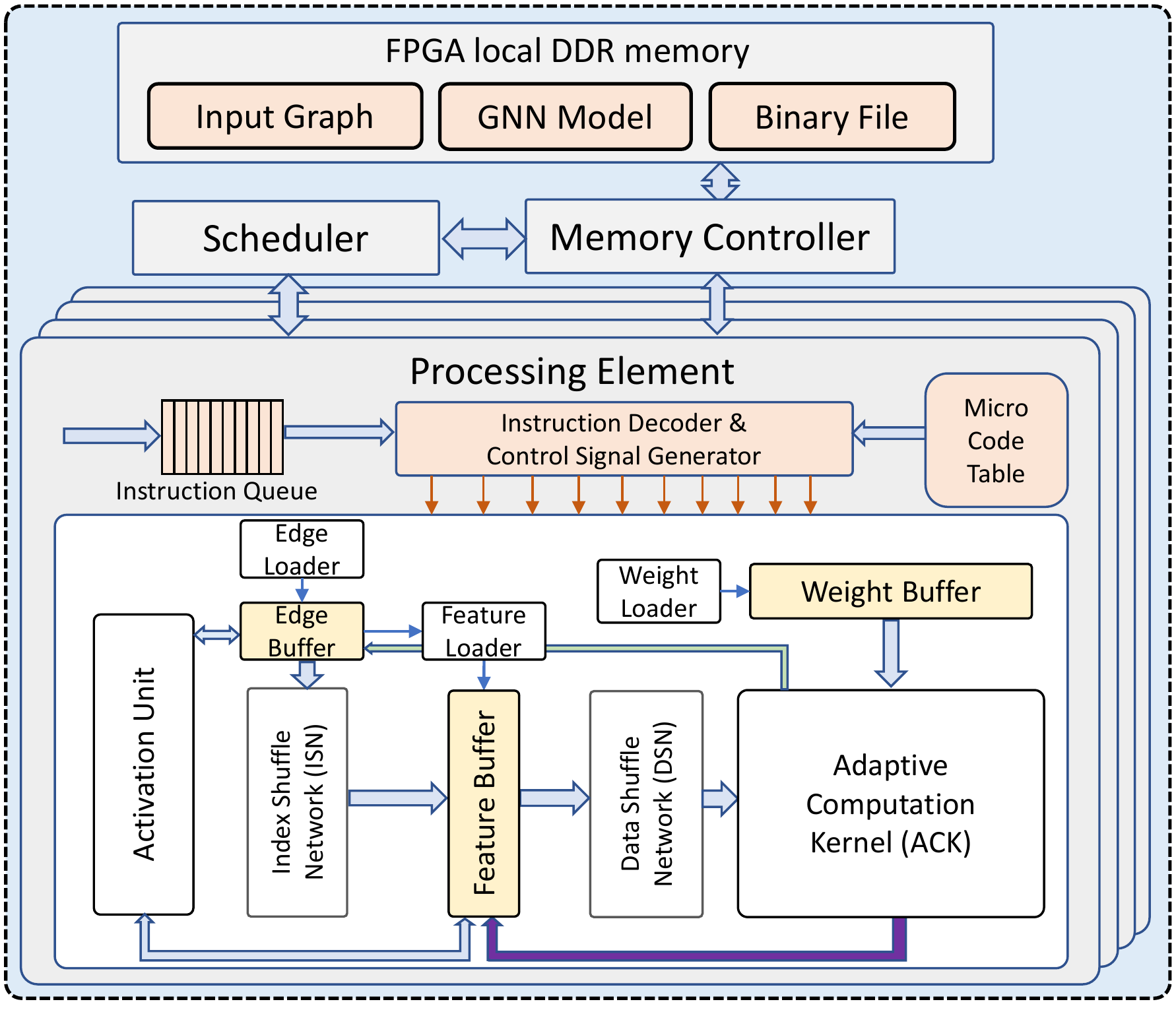}
     \caption{Hardware architecture}
     \label{fig:architecuture-high-level}
\end{figure}
Figure \ref{fig:architecuture-high-level} depicts the proposed hardware architecture. There are $N_{\text{pe}}$ Processing Elements (PEs) working in parallel. At runtime, the Scheduler reads the executable/binary file from the FPGA DDR and assigns the workload to PEs (see  Section \ref{subsec:Kernel-Mapping-task-scheduling}). Each PE has an Instruction Queue (IQ) to receive the incoming instructions assigned by the Scheduler. The Instruction Decoder \& Control Signal Generator reads the instructions from IQ and generates the control signals for the hardware modules. Each PE has a Weight Buffer to store the weight matrices, an Edge Buffer to store the edges, and a Feature Buffer to store the vertex feature vectors. Each buffer has a data loader\&writer that communicates with the FPGA DDR. Each PE has an Adaptive Computation Kernel, which is the key novelty in our hardware design. The Adaptive Computation Kernel (ACK, Figure \ref{fig:arch-ack}) can execute various computation kernels of GNNs.

\vspace{0.1cm}
\noindent \textbf{Hardware parameters}: The proposed architecture is defined by the following hardware parameters: (1) the number of Processing Elements $N_{\text{pe}}$, (2) the dimension of the Adaptive Computation Kernel (ACK) $P_{\text{sys}}\times P_{\text{sys}}$,  (3) dimensions of buffers, including the dimension of Weight Buffer $N_{W}\times P_{\text{sys}}$, the dimension of Edge Buffer $N_{E}\times 3$, the dimension of Feature Buffer $N_{F1}\times N_{F2}$, (4) the set of arithmetic operations supported by the ACK and the Activation Unit.

\section{Microarchitecture}
\label{Sec:Microarchitecture}

{ In this section, we first introduce the data format used by GraphAGILE (Section \ref{subsec:data-format}) and then discuss the key computation kernels in GNNs (Section \ref{subsec:computation-kernels}). Next, we describe the proposed instruction set of GraphAGILE (Section \ref{subsec:instruction-set}), and provide details of the microarchitecture that supports the proposed instruction set, including the datapath (Section \ref{subsec:Adaptive-Computation-Kernel}) and the on-chip memory organization (Section \ref{subsec:on-chip-memory}).}

\subsection{Graph Data Format}
\label{subsec:data-format}

 We use $\bm{h}_{i}^{l}$ to denote the \emph{feature vector} of vertex $v_{i}$ at layer $l$ (Table \ref{tab:notations}). We use the Coordinate Format (COO) to capture all graph \emph{edges}.  Each edge is a 3-tuple $(src, dst, weight)$ denoting the source vertex index, destination vertex index, and edge weight, respectively. We construct the feature matrix $\bm{H}$ by stacking feature vectors. Each row of $\bm{H}$ is the feature vector of a vertex. Denote $\bm{A}$ as the sparse adjacency matrix where for an edge $(u, v, w)$, we have $\bm{A}_{u,v} = w$.

\subsection{Computation Kernels in GNNs}
\label{subsec:computation-kernels}

We identify the following key computation kernels:

\vspace{0.1cm}
\noindent \textbf{General dense-dense matrix multiplication (GEMM)}: $\phi()$,  $\psi()$ and  $\rho()$ can involve GEMM, where the feature matrix $\bm{H}$ is multiplied by weight matrix $\bm{W}$. For example, in the $\phi()$ of GAT \cite{velivckovic2017graph}, $\bm{H}$ is multiplied by  $\bm{W}_{\text{att}}$ for calculating edge weights (Equation \ref{eq:gat-attention}).  In $\psi()$  and $\rho()$ of GraphSAGE \cite{hamilton2017inductive},   $\bm{H}$ is multiplied by  $\bm{W}$ to obtain the updated feature vectors. In general, $\bm{H}$ is a large dense matrix with height equal to $|\mathcal{V}|$ and  $\bm{W}$ is a small dense matrix (e.g., $\bm{W}$ has size $256\times 256$ in \cite{hamilton2017inductive}).


\vspace{0.1cm}
\noindent \textbf{Sparse-dense matrix multiplication (SpDMM)}: According to Equations \ref{eq:edge-ise} and \ref{eq:node-ise}, the vertices propagate the messages $\bm{m}_{e}^{l+1}$ along the outgoing edges. Then each vertex aggregates the incoming messages through  $\rho()$. The above message passing process is equivalent to SpDMM $\bm{A} \cdot \bm{H}$. 

\vspace{0.1cm}
\noindent \textbf{Sampled dense-dense matrix multiplication (SDDMM)}: According to \cite{fey2019fast}, in edge-wise computation (Equation \ref{eq:edge-ise}), many GNN models calculate edge weights using the dot product of the feature vectors of the source and destination vertices. The above computation process corresponds to SDDMM operation $\bm{A} \odot (\bm{H}\bm{H}^{T})$, where $\odot$ is the element-wise multiplication. \emph{Sampled} means that the required results are sampled from  $(\bm{H}\bm{H}^{T})$ based on the non-zero elements in $\bm{A}$. For each non-zero element $\bm{A}_{i,j}$. we calculate $\bm{A}_{i,j} = \langle \bm{H}_{i}, \bm{H}_{j} \rangle$. Therefore, the basic operation in SDDMM is the vector inner product.

\vspace{0.1cm}
\noindent \textbf{Other computation kernels}: GNNs also involve vector addition (e.g., residual connection), element-wise activation (e.g., ReLU, Softmax), batch normalization.

\subsection{Instruction Set}
\label{subsec:instruction-set}

\begin{figure*}[h]
     \centering
     \includegraphics[width=17cm]{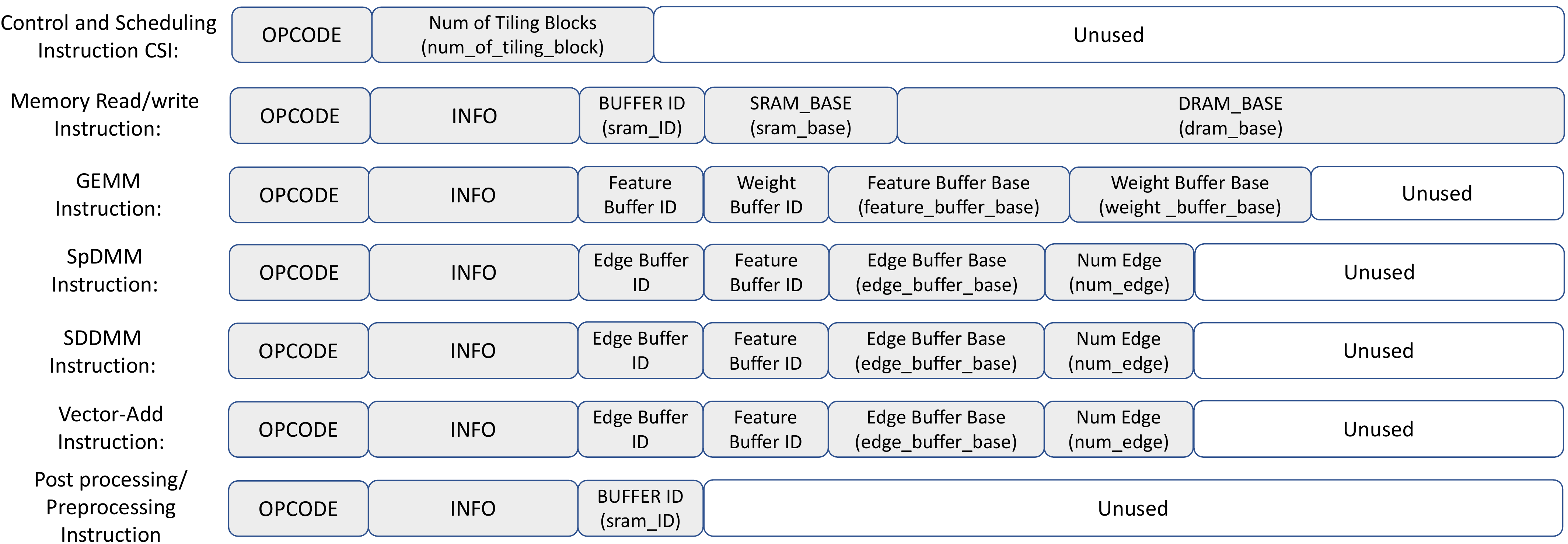}
     \caption{GraphAGILE high-level instruction fields}
     \label{app:fig:instruction-sets}
\end{figure*}

\subsubsection{High-level Instructions} The proposed instruction set is comprised of \emph{high-level instructions} and \emph{microcode}. All the high-level instructions have a uniform 128-bit length, and the instruction fields are depicted in Figure \ref{app:fig:instruction-sets}. The $\verb|OPCODE|$ field indicates the type of instruction. Other fields contain instruction-specific information. 
\begin{itemize}
    \item \emph{Control and Scheduling Instruction (CSI)}: A CSI contains the meta data of  a computation layer in the intermediate representation (Section \ref{subsec:IR}). Based on the CSI, the scheduler assigns the workloads of a layer to the PEs.
    \item \emph{Memory Read/Write Instruction}: A memory read/write instruction initiates  data communication (model weights, edges, vertex feature vectors) with FPGA DDR memory. 
    \item \emph{GEMM Instruction}: A GEMM instruction contains the information (e.g., matrix size, buffer ID that stores the matrices) of the matrix multiplication between the weight matrix (in the Weight Buffer) and feature matrix (in the Feature Buffer).
    \item \emph{SpDMM Instruction}: A SpDMM instruction performs multiplication of $\bm{A}$ and $\bm{H}$. The instruction specifies the number of non-zero elements in $\bm{A}$ (which enables edge-centric computation of SpDMM. See Section \ref{subsec:Adaptive-Computation-Kernel}) and buffer ID that stores $\bm{A}$. 
    \item \emph{SDDMM Instruction}: Similar to the SpMM instruction, it specifies the number of non-zero elements in $\bm{A}$ and the buffer IDs that store $\bm{A}$ and $\bm{H}$. 
    \item \emph{Other instructions}: There are other instructions including the Initialization Instruction, Activation Instruction, etc.
\end{itemize}

\subsubsection{Microcode} A high-level instruction defines a computation task in coarse-grained granularity. To execute a high-level instruction, the Instruction Decoder \& Control Signal Generator translates it to a sequence of microcode that has fine-grained granularity that can be executed by ACK.   The translation is through looking up the Microcode Table. For example, a GEMM instruction defines the multiplication of a large feature matrix (stored in Feature Buffer) and a large weight matrix (stored in Weight Buffer).
The GEMM instruction is decomposed into block matrix multiplication (BlockMM), where block size corresponds to the dimension of ACK.
The microcodes of GEMM use a three-level nested loop to execute the BlockMM on ACK. The microcodes of GEMM, SpDMM, {and} SDDMM are described as follows:

\vspace{0.1cm}
\noindent \textbf{Microcode of GEMM instructions}:
\begin{figure}[h]
     \centering
     \vspace{-0.3cm}
     \includegraphics[width=7cm]{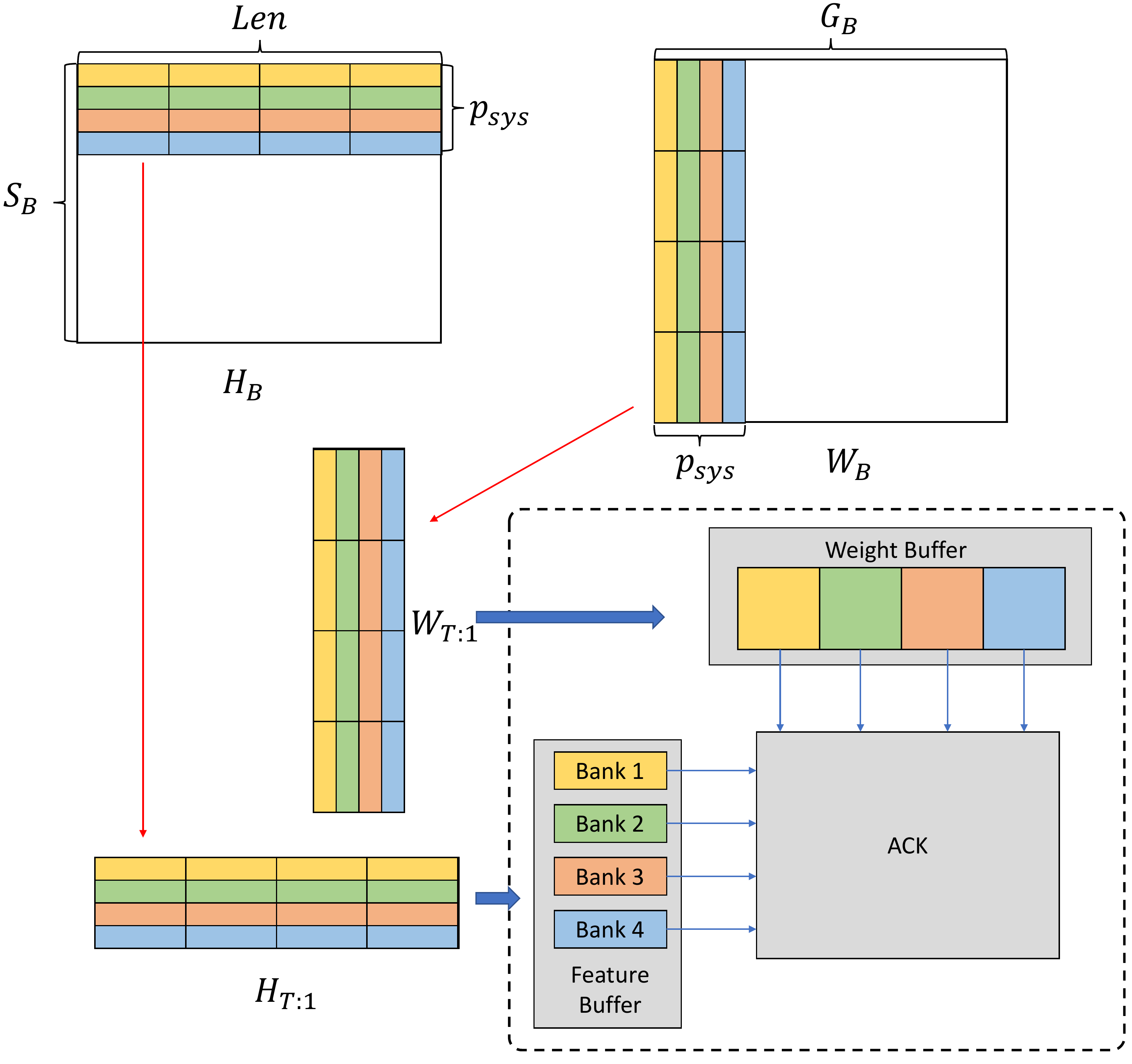}
     \vspace{-0.2cm}
     \caption{GEMM between a block of feature matrix $\bm{H}_{B}$ (stored in Feature Buffer) and a block of weight matrix $\bm{W}_{B}$ (stored in Weight Buffer)}
     \label{fig:structure-ALU}
\end{figure}
A high-level GEMM instruction is translated to a sequence of microcode to execute the GEMM between a block of feature matrix $\bm{H}_{B}\in \mathbb{R}^{S_{B}\times Len}$ and a block of weight matrix $\bm{W}_{B} \in \mathbb{R}^{Len \times G_{B}}$. The Pseudocode of the sequence of microcode is described in Algorithm \ref{algo:Pseudocode-of-GEMM-instruction}. In GEMM mode, the Adaptive Computation Kernel (ACK) works as a 2-D systolic array of size $p_{\text{sys}} \times p_{\text{sys}}$ using output-stationary dataflow. In each clock cycle, ACK receives  $p_{\text{sys}}$ data from Feature Buffer and $p_{\text{sys}}$ data from Weight Buffer, respectively. In Feature Buffer, $\bm{H}_{B}$ is further partitioned to small data tiles along row dimensions, and each data tile $\bm{H}_{T:i}$ has $p_{\text{sys}}$ rows. Similarly, in Weight Buffer, $\bm{W}_{B}$ is partitioned to small data tiles along {column} dimension and each data tile $\bm{W}_{T:i}$ has $p_{\text{sys}}$ columns of $\bm{W}_{B}$. $\bm{H}_{\text{out}:ij}$ denotes the result of the multiplication between $\bm{H}_{T:i}$ and $\bm{W}_{T:j}$.

\begin{algorithm}
 \caption{Pseudocode of GEMM microcode}
 \begin{algorithmic}[1]
 \renewcommand{\algorithmicrequire}{\textbf{Input:}}
\renewcommand{\algorithmicensure}{\textbf{Output:}}
 \Require $\bm{H}_{B}$; $\bm{W}_{B}$
 \Ensure $\bm{H}_{\text{out}}$
\For{$i \leftarrow 1$ to $\frac{S_{B}}{p_{\text{sys}}}$}
    \For{$j \leftarrow 1$ to $\frac{G_{B}}{p_{\text{sys}}}$}
        \State // {\color{blue}Pipelined execution of $\bm{H}_{\text{out}:ij} = \bm{H}_{T:i} \times \bm{W}_{T:j}$}
        \For{$k \leftarrow 1$ to $Len$ \textbf{Parallel}}
            \State Load the $p_{sys}$ data of $k^{\text{th}}$ column of $\bm{H}_{T:i}$ \hspace*{5em}  and send them to ACK
            \State Load the $p_{sys}$ data of  $k^{\text{th}}$ row of $\bm{W}_{T:j}$  and  \hspace*{5em}  send them to ACK
        \EndFor
    \EndFor
\EndFor
  \end{algorithmic} 
 \label{algo:Pseudocode-of-GEMM-instruction}
 \end{algorithm}

\begin{algorithm}
 \caption{Pseudocode of SpDMM microcode}
 \begin{algorithmic}[1]
 \renewcommand{\algorithmicrequire}{\textbf{Input:}}
\renewcommand{\algorithmicensure}{\textbf{Output:}}
 \Require $\bm{H}_{B}$; $\bm{A}_{B}$; number of edges in $\bm{A}_{B}$: $N_{e}$
 \Ensure $\bm{H}_{\text{out}}$
\For{$i \leftarrow 1$ to $\frac{2N_{e}}{p_{\text{sys}}}$}  {\color{blue}\Comment{Pipelined execution of SpDMM}}
    \State Load $\frac{p_{\text{sys}}}{2}$ unprocessed edges from $\bm{A}_{B}$ in Edge Buffer 
    \State Send the $\frac{p_{\text{sys}}}{2}$ edges to Index Shuffle Network (ISN)
\EndFor
  \end{algorithmic} 
 \label{algo:Pseudocode-of-SpDMM-instruction}
 \end{algorithm}
 
\vspace{0.1cm}
\noindent \textbf{Microcode of SpDMM instructions}:
A high-level SpDMM instruction is translated to a sequence of microcode to execute the SpDMM between a block of feature matrix $\bm{H}_{B}$ (stored in the Feature Buffer) and a block of sparse adjacency matrix $\bm{A}_{B}$ (stored in the Edge Buffer). The execution of SpDMM is edge-centric (See Section \ref{subsec:Adaptive-Computation-Kernel}). Therefore, in each clock cycle, $\frac{p_{\text{sys}}}{2}$ unprocessed edges in $\bm{A}_{B}$ are fetched from Edge Buffer. The $\frac{p_{\text{sys}}}{2}$ edges are sent to Index Shuffle Network to execute feature aggregation.

\begin{algorithm}
 \caption{Pseudocode of SDDMM microcode}
 \begin{algorithmic}[1]
 \renewcommand{\algorithmicrequire}{\textbf{Input:}}
\renewcommand{\algorithmicensure}{\textbf{Output:}}
 \Require $\bm{H}_{B}$; $\bm{A}_{B}$; number of edges in $\bm{A}_{B}$: $N_{e}$
 \Ensure weights of all the edges in $\bm{A}_{B}$
\For{$i \leftarrow 1$ to $\frac{2N_{e}}{p_{\text{sys}}}$} {\color{blue}\Comment{Pipelined execution of SDDMM}}
    \State Load $\frac{p_{\text{sys}}}{2}$ unprocessed edges from $\bm{A}_{B}$ in Edge Buffer
    \State Extract the $\frac{p_{\text{sys}}}{2}$ $src$ indices and $\frac{p_{\text{sys}}}{2}$ $dst$ indices
    \State Send the $p_{\text{sys}}$ indices to ISN
\EndFor
  \end{algorithmic} 
 \label{algo:Pseudocode-of-SDDMM-instruction}
 \end{algorithm}
\vspace{0.1cm}
\noindent \textbf{Microcode of SDDMM instructions}:
A high-level SDDMM instruction is translated to a sequence of microcode to execute the SDDMM using a block of feature matrix $\bm{H}_{B}$ (stored in the Feature Buffer) and a block of sparse adjacency matrix $\bm{A}_{B}$ (stored in the Edge Buffer). Similar to SpDMM, the execution of SDDMM is edge-centric (See Section \ref{subsec:Adaptive-Computation-Kernel}). In each clock cycle, $\frac{p_{\text{sys}}}{2}$ unprocessed edges in $\bm{A}_{B}$ are fetched from Edge Buffer. The $\frac{p_{\text{sys}}}{2}$ $src$ indices and $\frac{p_{\text{sys}}}{2}$ $dst$ indices  are extracted from the $\frac{p_{\text{sys}}}{2}$ unprocessed edges. Then, the total $p_{\text{sys}}$ indices are sent to Index Shuffle Network (ISN) to execute the SDDMM of $\bm{A}_{B}$ and $\bm{H}_{B}$.

\subsection{Various Execution Modes}
\label{subsec:Adaptive-Computation-Kernel}
\begin{figure}[h]
     \centering
     \includegraphics[width=8.5cm]{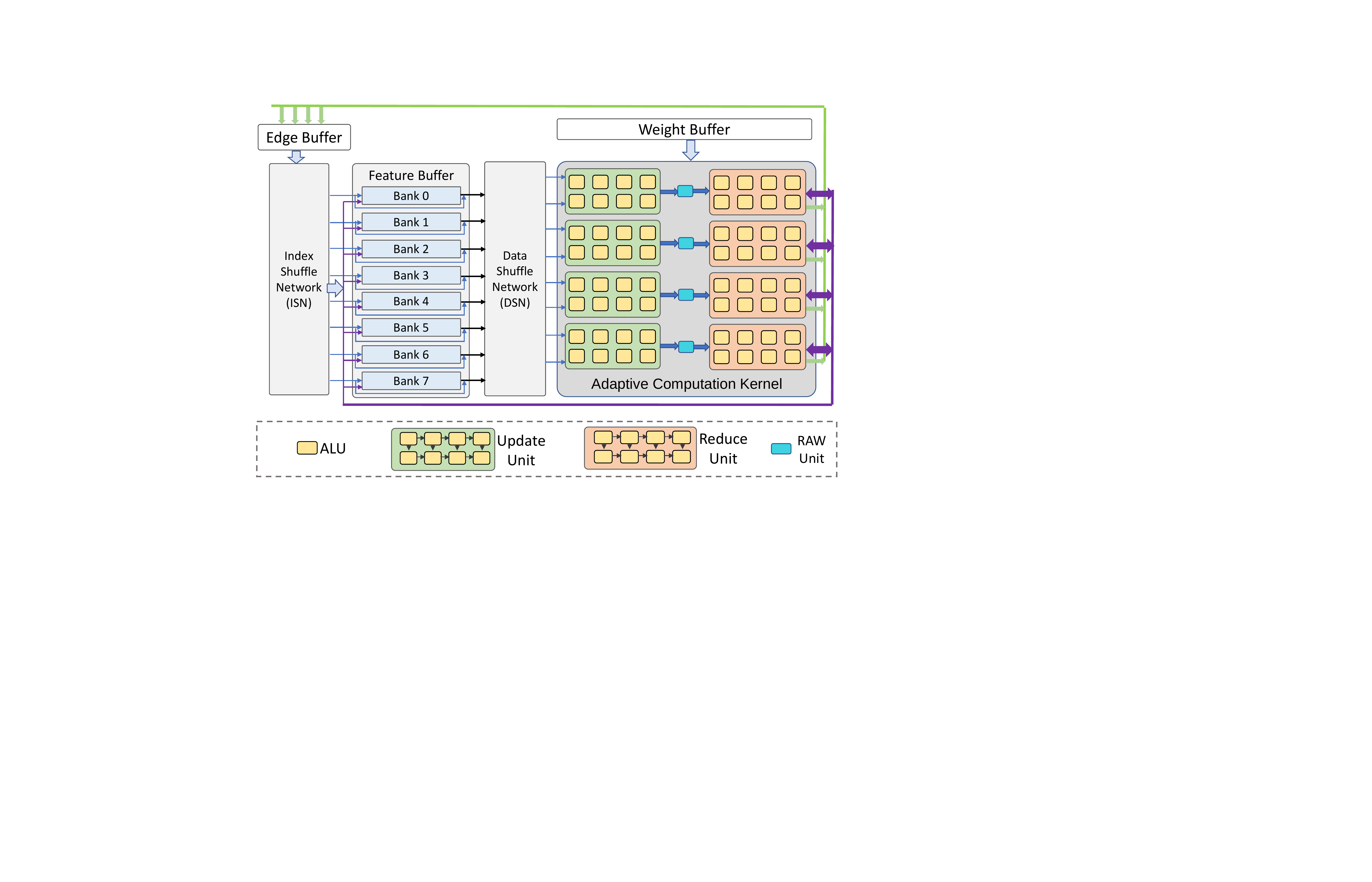}
     \caption{Adaptive Computation Kernel (when $p_{\text{sys}} = 8$) with ISN and DSN. The interconnections among ALUs are specified in Figure \ref{fig:three-execution-mode}. }
     \label{fig:arch-ack}
\end{figure}

\begin{figure*}[h]
     \centering
     \includegraphics[width=18cm]{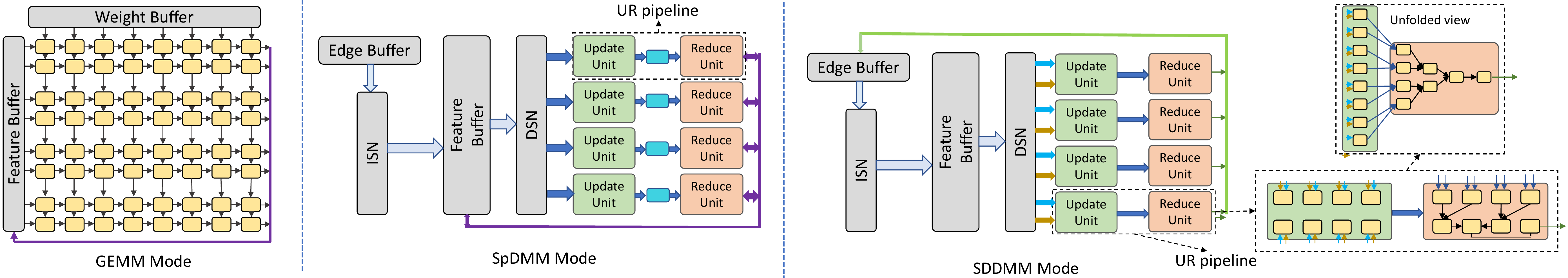} 
     \caption{The datapath of GEMM mode, SpDMM mode, SDDMM mode}
     \label{fig:three-execution-mode}
 \end{figure*}

 As shown in Figure \ref{fig:arch-ack},  an ACK contains an array of Arithmetic Logic Units (ALUs) of size $p_{\text{sys}}\times p_{\text{sys}}$, where $p_{\text{sys}}$ is the power of 2. An ALU can execute various arithmetic operations, including Multiplication, Addition, Accumulation, Min, Max, etc. The interconnections among ALUs are shown in Figure \ref{fig:three-execution-mode}. The array of ALUs { is} divided into Update Units and Reduce Units. An Update Unit or a Reduce Unit has size $(p_{\text{sys}}/2) \times 2$. 
 The Feature Buffer has $p_{\text{sys}}$ parallel memory banks.  $\bm{h}_{i}$ is stored in bank $(i\mod p_{\text{sys}})$. There are two interconnection networks -- Index Shuffle Network (ISN) and Data Shuffle Network (DSN). The ISN routes edges to the memory banks of Feature Buffer for fetching the features of incident vertices. The DSN routes the input data (vertex features with the edge) to  Adaptive Computation Kernel. The routing is based on the least significant $\log(p_{\text{sys}})$ bits of the vertex index. 
The ACK has various execution modes, including \emph{GEMM mode}, \emph{SpDMM mode}, \emph{SDDMM mode}, and \emph{Vector-Addition mode}.  Each ALU maintains multiplexers with control logic to select the input and output ports for an execution mode. The mode switching incurs the overhead of only one clock cycle.

\vspace{0.1cm}
\noindent \textbf{GEMM mode}:  The array of ALUs is organized as a two-dimensional systolic array with fully localized interconnection. GEMM mode supports dense matrix multiplication of feature matrix $\bm{H}$ and weight matrix $\bm{W}$. Weight Buffer streams the weight matrix into the systolic array, and Feature Buffer streams multiple vertex feature vectors into the systolic array. Systolic array of size $p_{\text{sys}}\times p_{\text{sys}}$ executes  $p_{\text{sys}}^{2}$ Multiply-Accumulation operations per clock cycle.


\begin{algorithm}
\caption{SpDMM following Scatter-Gather paradigm}\label{alg:scatter-gather}
\begin{small}
\begin{algorithmic}

\While {not done}
\For{each edge $e(src,~dst,~weight)$ }  {\color{blue}\Comment{Scatter Phase}}
\State Fetch $src.features$ from Feature Buffer
\State Form input pair ($src.features$, $e$) 
\EndFor

\For{each input pair}
{\color{blue}\Comment{Gather Phase}}
\State Produce $u \gets ${Update($src.features,~ e.weight$)}
\State {Update $v_{dst} \gets$ {Reduce($u.features$)}}
\EndFor

\EndWhile
\end{algorithmic}
\end{small}
\end{algorithm}

\noindent \textbf{SpDMM mode}: As shown in Algorithm \ref{alg:scatter-gather}, SpDMM is executed following the Scatter-Gather paradigm.  The array of ALUs in ACK is divided into multiple Update Units and Reduce Units. In each Update Unit, the ALUs are organized as a vector multiplier that multiplies the vertex feature vector by the edge weight. In each  Reduce Unit, the ALUs execute the element-wise reduction operation $\rho()$. Suppose a vertex is defined by $(src, features)$, where $src$ denotes the source vertex index and the $features$ is the feature vector of the source vertex.  
The generated intermediate results by the Update Units are represented by $(dst, features)$. 
The intermediate results are applied to the destination vertex $v_{dst}$ by the Reduce Unit. 
An Update Unit and a Reduce Unit form an ``UR-pipeline''. The computation of SpDMM is driven by unprocessed edges (i.e., \emph{edge-centric processing} \cite{zhou2019hitgraph}). Unprocessed edges are fetched from Edge Buffer to ISN.
In ISN, an edge $e$ is routed to the corresponding memory bank in Feature Buffer to fetch $src.features$, thus forming the input pair $(src.features, e)$. Then, the DSN routes the input pairs to the UR pipelines based on the $dst$ of the edge.
The input pairs having $e.dst=i\times p_{\text{sys}} +k$ ($0 \leqslant k < p_{\text{sys}} $) will be routed to the ${\lfloor k/2 \rfloor}^{\text{th}}$ UR pipeline. This is because the output port of ${\lfloor k/2 \rfloor}^{\text{th}}$ UR pipeline is connected to bank $\lfloor k/2 \rfloor$ and bank $\lfloor k/2 \rfloor + 1$ of Feature Buffer, where $v_{e.dst}$ is stored. Then, the UR pipeline processes the input pair, and the intermediate result generated by the input pair is applied to the destination vertex $v_{e.dst}$.
$p_{\text{sys}}/2$ input pairs can be processed by the $p_{\text{sys}}/2$ UR pipelines  concurrently.



\vspace{0.1cm}
\noindent \textbf{SDDMM mode}: The basic operation is the inner product of two feature vectors. For each edge $(src, dst)$, the feature vectors $\bm{h}_{src}$ and $\bm{h}_{dst}$ are fetched from the Feature Buffer. The result of the inner product of $\bm{h}_{src}$ and $\bm{h}_{dst}$ becomes the weight of the edge  $(src, dst)$. To support the inner-product, the ALUs in a UR pipeline form a multiply-adder tree. The topological structure of the multiply-adder tree is shown in Figure \ref{fig:three-execution-mode}. Similar to SpDMM, the execution of SDDMM is edge-centric. For an edge $(src, dst)$, $src$ and $dst$ are routed to the corresponding memory banks of Feature Buffer to fetch $\bm{h}_{src}$ and $\bm{h}_{dst}$. The inner product of $\bm{h}_{src}$ and $\bm{h}_{dst}$ is executed by a UR pipeline. The ACK can execute $p_{\text{sys}}/2$ vector inner products of length $p_{\text{sys}}$ during each clock cycle. The dot product of two feature vectors of length $|\bm{h}_{i}|$ is executed in $\left\lceil \frac{|\bm{h}_{i}|}{p_{\text{sys}}} \right\rceil$ cycles and the intermediate result is stored at the root node of the adder tree for accumulation.

\vspace{0.1cm}
 \noindent \textbf{Vector Addition Mode}: In Vector Addition Mode, the basic operation is the addition of two feature vectors. An Update Unit (See Figure \ref{fig:arch-ack}) works as a vector adder. To add $\bm{h}_{\text{u}}$ and $\bm{h}_{\text{v}}$, the indices $u$ and $v$ are routed through Index Shuffle Network (ISN) to Feature Buffer to fetch $\bm{h}_{\text{u}}$ and $\bm{h}_{\text{v}}$. Then $\bm{h}_{\text{u}}$ and $\bm{h}_{\text{v}}$ are routed to an Update Unit through Data Shuffle Network (DSN) to perform vector addition. The results will bypass the Reduction Unit and are sent back to Feature Buffer. The ACK can execute $p_{\text{sys}}/2$ vector additions of length $p_{\text{sys}}$ at each clock cycle. Two feature vectors of length $|\bm{h}_{i}|$ can be added in $\lceil \frac{|\bm{h}_{i}|}{p_{\text{sys}}} \rceil$ cycles.



\subsection{Parallel On-chip Memory Access}
\label{subsec:on-chip-memory}


The Feature Buffer supports parallel memory access patterns of various computation modes enabled by ISN and DSN. Feature Buffer has $p_{\text{sys}}$ parallel memory banks, and the feature vector of vertex $v_{i}$ is stored in bank $(i\mod p_{\text{sys}})$. Edge Buffer can output $p_\text{sys}$ edges at each clock cycle by having port width $p_\text{sys}d_{e}$, where $d_{e}$ is the bit width of an edge. ISN performs all-to-all interconnection between Edge Buffer and Feature Buffer. DSN performs all-to-all interconnection between Feature Buffer and ACK. The ISN and  the DSN are implemented using the  butterfly network \cite{choi2021hbm}

\vspace{0.1cm}
\noindent \textbf{Parallel memory accesses in GEMM mode}: The ACK directly fetches $p_{\text{sys}}$ features from $p_{\text{sys}}$ memory banks per clock cycle. No data shuffling is required for GEMM. The Weight Buffer also has $p_{\text{sys}}$ memory banks that can output $p_{\text{sys}}$  data of the weight matrix each clock cycle. 

\vspace{0.1cm}
\noindent \textbf{Parallel memory accesses in SpDMM mode}: $p_{\text{sys}}/2$ edges $\{e_{1},e_{2},...,e_{p_{\text{sys}/2}} \}$ are sent to ISN simultaneously. The edges are routed to the corresponding memory banks of Feature Buffer based on their $src$.  $p_{\text{sys}}/2$ edges will generate $p_{\text{sys}}/2$  input pairs $(src.features,e)$ after fetching the feature vectors. Then the $p_{\text{sys}}/2$ input pairs are routed to the corresponding UR pipelines based on their $e.dst$.

\vspace{0.1cm}
\noindent \textbf{Parallel memory accesses in SDDMM mode}: $p_{\text{sys}}/2$ edges $\{e_{1},e_{2},...,e_{p_{\text{sys}}/2} \}$ are fetched from the Edge Buffer in each cycle.   The $p_{\text{sys}}/2$ $src$ indices and $p_{\text{sys}}/2$ $dst$ indices $\{src_{1}, dst_{1}, src_{2}, dst_{2},..., src_{p_{\text{sys}}/2}, dst_{p_{\text{sys}}/2}\}$ of $p_{\text{sys}}/2$ edges are sent to $p_{\text{sys}}$ input ports of   ISN. The ISN routes the $p_{\text{sys}}$ indices to the Feature Buffer to fetch the $p_{\text{sys}}$ vertex feature vectors from the Feature Buffer. Then, the  $p_{\text{sys}}$  feature vectors are routed to the $p_{\text{sys}}/2$ UR pipelines of ACK.
The $i^{\text{th}}$ UR pipeline  performs the inner product of $\bm{h}_{src_{i}}$ and $\bm{h}_{dst_{i}}$.

\section{Compiler}
\label{sec:compiler}

We develop a compiler that reads the user-defined GNN model and input graph, and generates a sequence of instructions. User defines the  GNN model using the high-level API in Pytorch Geometric (PyG) Library \cite{fey2019fast}, which is a general framework for GNNs. There are two phases for instruction generation -- \emph{translation phase} and \emph{optimization phase}. In the translation phase,  the Input Parser generates the Intermediate Representation (IR) from the inputs. In the optimization phase,  the compiler performs four-step optimizations and generates the output instruction sequence: (1) \textbf{Step 1}: the compiler reorders the computation graph based on the theoretical computation complexity. (2) \textbf{Step 2}:  the compiler merges some adjacent layers to communicate intermediate data through on-chip memory. (3) \textbf{Step 3}:  the compiler performs data partitioning based on the available on-chip memory to optimize off-chip data communication and enable dynamic task scheduling, (4) \textbf{Step 4}: the compiler maps various kernels to ACK, and performs task scheduling to hide the data communication overhead and achieve dynamic load balance.

\begin{small}
\begin{lstlisting}[language=Python, caption=An user-defined GNN model using PyG \cite{fey2019fast}, label={lst:GNN-example}]
import torch
from torch import Tensor
from torch_geometric.nn import GCNConv
from torch_geometric.datasets import Planetoid

dataset = Planetoid(root='.', name='Cora')

class GCN(torch.nn.Module):
    def __init__(self, in_ch, hidden_ch, out_ch):
        super().__init__()
        self.conv1 = GCNConv(in_ch, hidden_ch)
        self.conv2 = GCNConv(hidden_ch, out_ch)
    def forward(self, x:Tensor, edge_index: Tensor):
        x = self.conv1(x, edge_index).relu()
        x = self.conv2(x, edge_index)
        return x
        
model = GCN(dataset.num_features, 16, dataset.num_classes)  
\end{lstlisting}
\end{small}

\subsection{Intermediate Representation}
\label{subsec:IR}

\begin{table}[ht]
\centering
\caption{IR of a computation layer}
\begin{tabular}{|l|l|}
\hline
\textbf{Layer Type} & \begin{tabular}[|c|]{@{}c@{}} Aggregate(0), Linear(1), \\ Vector-Inner(2), Vector-Add(3), \\ Activation(4), BatchNorm(5) \end{tabular} \\ 
\hline
\textbf{Layer ID}  & 1,2,3,... \\
\hline
\textbf{Parent Layer IDs}  &  Parent1\_ID, ... \\
\hline
\textbf{Child Layer IDs}  &  Child1\_ID, ... \\
\hline
\textbf{Input Dimension}  & $f_{\text{in}}$  \\
\hline
\textbf{Output Dimension}  & $f_{\text{out}}$ \\
\hline
\textbf{\# of vertices} & $|\mathcal{V}|$  \\
\hline
\textbf{\# of edges} & $|\mathcal{E}|$  \\
\hline
\textbf{Aggregation operator} &  Max, Sum, Min, Mean \\
\hline
\textbf{Activation type} & ReLU, PReLU, Swish, Exp\\
\hline
\textbf{Activation enabled} & True, False\\
\hline
\end{tabular}
\label{tab:inter-representation}
\end{table}

\begin{figure}[h]
     \centering
     \includegraphics[width=7.5cm]{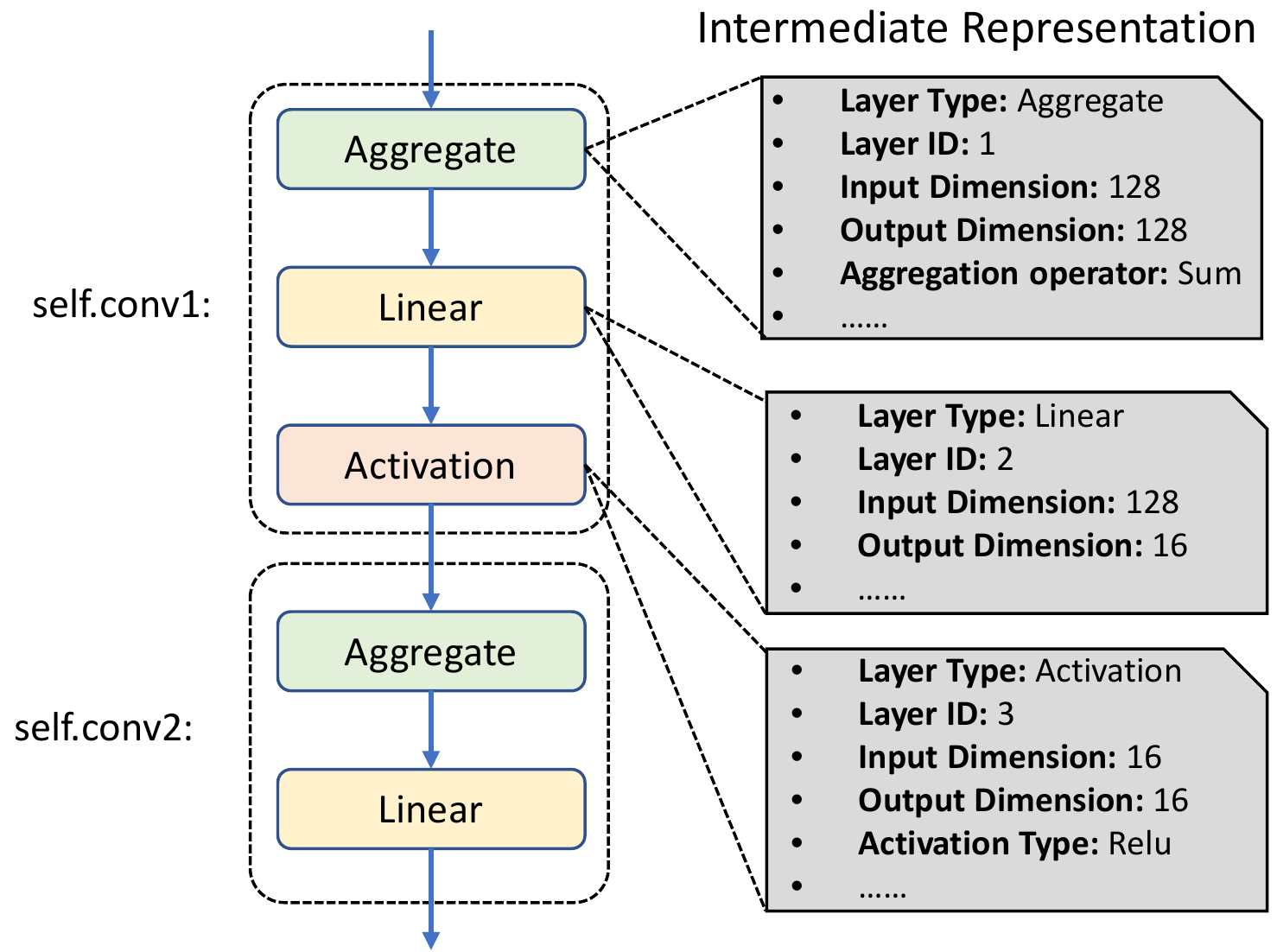}
     \caption{The computation graph of the GNN in Listing \ref{lst:GNN-example}}
     \label{fig:IR-example}
\end{figure}

We define a unified Intermediate Representation (IR) for each type of computation layer (Table \ref{tab:inter-representation}). A GNN layer can be decomposed into a sequence of computation layers. We identify six types of computation layers -- \emph{Aggregate}, \emph{Linear}, \emph{Vector-Inner}, \emph{Vector-Add}, \emph{Activation} and \emph{BatchNorm}. The six types of layers can represent a broad range of models because (1) the key computation kernels of GNNs (SpDMM, GEMM, {and} SDDMM) can be represented as \emph{Aggregate}, \emph{Linear}, or \emph{Vector-Inner}, (2) the auxiliary kernels such as non-linear activation, residual connection, batch normalization can be represented using others lightweight layers (e.g., \emph{Vector-Add}, \emph{Activation}, and \emph{BatchNorm}). The compiler translates the GNN model to a computation graph, with each node being the IR of a layer. For example, the GNN model \cite{kipf2016semi} in Listing \ref{lst:GNN-example} is translated to the computation graph in Figure \ref{fig:IR-example}. The abstraction of each type of computation layer is described in the following:

\vspace{0.1cm}
\noindent \underline{\textbf{Aggregate layer}}: The inputs are the vertex feature vectors  $\{\bm{h}_{i}^{l-1} \in \mathbb{R}^{f_{\text{in}}}:v_{i} \in \mathcal{V}\}$ and the edges $\{e:e\in \mathcal{E}\}$. The output feature vector of each vertex is calculated by:
\begin{equation}
    \bm{h}_{i}^{l} = \text{AggOp}(\bm{A}_{j,i} \times \bm{h}_{j}^{l-1}, j \in \mathcal{N}(i)), \bm{h}_{i}^{l}\in \mathbb{R}^{f_{\text{out}}}
\end{equation}
where $f_{\text{in}}=f_{\text{out}}$ and $\text{AggOp}()$ is the element-wise Aggregation Operator defined in Table \ref{tab:inter-representation} (e.g., Max, Sum).  The Aggregate layer can be executed using SpDMM mode. 

\vspace{0.1cm}
\noindent \underline{\textbf{Linear layer}}: The inputs are the vertex feature vectors $\{\bm{h}_{i}^{l-1} \in \mathbb{R}^{f_{\text{in}}}:v_{i} \in \mathcal{V}\}$  and weight matrix $\bm{W}\in \mathbb{R}^{f_{\text{in}}\times f_{\text{out}}}$. The output feature vector of each vertex is calculated by:
\begin{equation}
\begin{split}
    \bm{H}_{\text{out}} & = [\bm{h}_{1}^{l};\bm{h}_{2}^{l};...; \bm{h}_{|\mathcal{V}|}^{l}]  = [\bm{h}_{1}^{l-1}\bm{W};\bm{h}_{2}^{l-1}\bm{W};...; \bm{h}_{|\mathcal{V}|}^{l-1}\bm{W}] \\
     & = [\bm{h}_{1}^{l-1};\bm{h}_{2}^{l-1};...; \bm{h}_{|\mathcal{V}|}^{l-1}]\bm{W} = \bm{H}_{\text{in}}\bm{W}
\end{split}
\end{equation}
where $[\bm{h}_{1}^{l-1};\bm{h}_{2}^{l-1};...; \bm{h}_{|\mathcal{V}|}^{l-1}]$ is the input feature matrix $\bm{H}_{\text{in}}$ and   $[\bm{h}_{1}^{l};\bm{h}_{2}^{l};...; \bm{h}_{|\mathcal{V}|}^{l}]$ is the output feature matrix $\bm{H}_{\text{out}}$. It can be executed using GEMM mode.

\vspace{0.1cm}
\noindent \underline{\textbf{Vector-Inner layer}}: The inputs are the vertex feature vectors  $\{\bm{h}_{i}^{l-1} \in \mathbb{R}^{f_{\text{in}}}:v_{i} \in \mathcal{V}\}$ and the edges  $e(i, j)$ without edge weight. The output is the weight of each edge calculated by:
\begin{equation}
    e( i, j).weight = \langle \bm{h}_{i}^{l-1},\bm{h}_{j}^{l-1} \rangle, \quad e(i, j) \in \mathcal{E}
\end{equation}

\vspace{0.1cm}
\noindent \underline{\textbf{Vector-Add layer}}: The Vector-Add layer adds feature vectors of two layers. This layer can be used to capture the residue connection design.

\vspace{0.1cm}
\noindent \underline{\textbf{Activation layer}}: The Activation layer applies the element-wise activation function (e.g., ReLU, PReLU, Swish, Exp) to vertex features or edge weights. 

\vspace{0.1cm}
\noindent \underline{\textbf{BatchNorm layer}}: The input is the feature vector of each vertex $\{\bm{h}_{i}^{l-1} \in \mathbb{R}^{f_{\text{in}}}:v_{i} \in \mathcal{V}\}$. A batch normalization operation \cite{ioffe2015batch} is applied to each vertex feature.

\subsection{IR Generation Workflow}
{ The proposed intermediate representation consists of two components: LayerIR and ModelIR. LayerIR is the IR of a computation layer that stores the parameters of a layer, as shown in Table \ref{tab:inter-representation}. ModelIR stores a list of LayerIRs and represents the computation graph corresponding to the target GNN model and the input graph. The implementation of LayerIR and ModelIR is demonstrated in Listing \ref{lst:IR-implementation}.}

{During compilation, the compiler first translates each computation layer into a LayerIR. Then, all the LayerIRs are connected to form a ModelIR, which represents the computation graph of the input GNN model and the input graph.  An example of the IR generation process for the GNN model in Listing \ref{lst:GNN-example} is illustrated in Listing \ref{lst:IR-generation} (Lines 12-39). Note that for illustration, the example in Listing \ref{lst:IR-generation} is an unfolded view of the IR generation process. In the actual implementation, the input parser automatically generates the ModelIR using a $\verb|for|$ loop. After IR generation, the compiler performs compiler optimizations, as shown in Listing \ref{lst:IR-generation} (Lines 42-46).}

\begin{small}
\begin{lstlisting}[language=Python, caption=The implementation of LayerIR and ModelIR, label={lst:IR-implementation}]
from collections import OrderedDict
## The IR of a computation layer
class LayerIR(object):
    def __init__(self):
        self._layertype = None  # Layer Tpe
        self._layerid = 0       # Layer ID
        self._parent_id = []    # Parent Layer IDs
        self._child_id = []     # Child Layer IDs
        self._fin = 0           # Input Dimension
        self._fout = 0          # Output Dimension
        self._nv = 0            # # of vertices
        self._ne = 0            # # of edges
        self._aggoperator=None  # AggOp()
        self._act = None        # Activation type
    def setparameter(self):
        # Setting the parameters for the computation layer 
    def complexity(self):
        # Return theoretical computation complexity of the computation layer

## The IR of a GNN model
class ModelIR(object):
    def __init__(self):
        self._layers= OrderedDict()
        self._graphs=None
        self._numl = 0
    def addlayers(self, layer):
        self._layers[layer._layerid] = layer
        self._numl = self._numl + 1
    def orderoptize(self):
        # Step 1: computation order optimization
    def layerfusion(self):
        # Step 2: layer fusion
    def datapartition(self):
        # Step 3: data partitioning
    def kernelMapping(self):
        # Step 4: kernel mapping
    def taskScheduling(self):
        # Step 4: task scheduling
    def codeGeneration(self):
        # Generating Instruction sequence

\end{lstlisting}
\end{small}

\begin{small}
\begin{lstlisting}[language=Python, caption=The example of IR generation, label={lst:IR-generation}]
dataset = 'Cora'
path = osp.join('.', 'data', dataset)
dataset = Planetoid(path, dataset, transform=T.NormalizeFeatures())
data = dataset[0]

nedges = data.edge_index.shape[1]
nvertices = data.x.shape[0]
nflen = data.x.shape[1]
edge_index = data.edge_index
edge_index = torch.transpose(edge_index, 0, 1)

## IR generation
GNN1 = ModelIR()
GNN1._graphs = data

aggregate1 = LayerIR()
aggregate1.setparameter(
    layertype = 'Aggregate', layerid = 1, parent_id = [], child_id = [2], fin = nflen, fout = nflen, nv = nvertices, ne=nedges, aggoperator=2, act=None, actenable=False, batchenable=False)
GNN1.addlayers(aggregate1)

linear1 = LayerIR()
linear1.setparameter(
    layertype = 'Linear', layerid = 2, parent_id = [1], child_id = [3], fin = nflen, fout = 16, nv = nvertices, ne=nedges, aggoperator=None, act=None, actenable=False, batchenable=False)
GNN1.addlayers(linear1)

activation1 = LayerIR()
activation1.setparameter(
    layertype = 'Activation', layerid = 3, parent_id = [2], child_id = [4], fin = 16, fout = 16, nv = nvertices, ne=nedges, aggoperator=None, act='ReLU', actenable=True, batchenable=False)
GNN1.addlayers(activation1)

aggregate2 = LayerIR()
aggregate2.setparameter(
    layertype = 'Aggregate', layerid = 4, parent_id = [3], child_id = [5], fin = 16, fout = 16, nv = nvertices, ne=nedges, aggoperator=2, act=None, actenable=False, batchenable=False)
GNN1.addlayers(aggregate2)

linear2 = LayerIR()
linear2.setparameter(
    layertype = 'Linear', layerid = 5, parent_id = [4], child_id = [], fin = 16, fout = 7, nv = nvertices, ne=nedges, aggoperator=None, act=None, actenable=False, batchenable=False)
GNN1.addlayers(linear2)

## IR optimizations
GNN1.orderoptize() # Step 1: computation order optimization
GNN1.layerfusion() # Step 2: layer fusion
GNN1.datapartition()  # Step 3: data partitioning
GNN1.kernelMapping()  # Step 4: kernel mapping
GNN1.taskScheduling() # Step 4: task scheduling
GNN1.codeGeneration('GNN1.ga')   # Generating Instruction sequence
\end{lstlisting}
\end{small}

\subsection{Computation Order Optimization}
\label{subsec:Computation-Order-Optimization}

We design the general rule for the  computation order optimization.  First, we define the  \emph{linear operator} in the aggregate layer:
\begin{definition}
In an Aggregate layer, the aggregation operator $\text{AggOp}()$ is a linear operator if $\text{AggOp}()$ satisfies the following two properties:
\begin{itemize}
    \item $\text{AggOp}(\bm{h}_{x} + \bm{h}_{y}) = \text{AggOp}(\bm{h}_{x}) + \text{AggOp}(\bm{h}_{y})$ for any $\bm{h}_{x} \in \mathbb{R}^{f}$ and  $\bm{h}_{y} \in \mathbb{R}^{f}$.
    \item $\text{AggOp}(c\bm{h}_{x}) = c\text{AggOp}(\bm{h}_{x})$ for any $\bm{h}_{x} \in \mathbb{R}^{f}$ and any constant $c$. 
\end{itemize}
\hspace{0.4cm} For example, Sum() is a linear operator while {Max()} is a non-linear operator. 
\end{definition}
Then, we identify the  exchangeability of computation order in Theorem \ref{theorem:computation-order}:
\begin{theorem}
\label{theorem:computation-order}
For a pair of adjacent Aggregate layer and Linear Layer, if the Aggregation operator $\text{AggOp}()$ of the Aggregate layer is a \emph{linear operator}, we can exchange the computation order of the Aggregate layer and Linear Layer.
\end{theorem} 
\begin{proof*}
The computation process of the adjacent Aggregate layer and Linear layer can be expressed as:
\begin{equation}
    \bm{h}_{i}^{l} = \text{AggOp}(\bm{A}_{j,i} \times \bm{h}_{j}^{l-1} \times \bm{W}, j \in \mathcal{N}(i))
\end{equation}
where AggOp() is the aggregation operator of the Aggregate layer and the $\bm{W}$ is the weight matrix of the Linear layer. Since the operator AggOp() is a linear operator, the above equation can be written as:
\begin{equation}
    \bm{h}_{i}^{l} = \text{AggOp}(\bm{A}_{j,i} \times \bm{h}_{j}^{l-1}, j \in \mathcal{N}(i))  \times \bm{W}
\end{equation}
Therefore, the computation order of this pair of Aggregate layer and Linear layer can be exchanged without affecting the final result. 
\end{proof*}
\vspace{0.2cm}
The computation order can affect the total computation complexity. The computation complexity (CC) of an Aggregate layer is:
\begin{equation}
    \text{CC}_{\text{Aggregate}} (f_{\text{in}}, f_{\text{out}}, |\mathcal{E}|)= 2 \cdot f_{in} \cdot |\mathcal{E}|, (f_{\text{in}} = f_{\text{out}})
\end{equation}
The computation complexity (CC) of a Linear layer is:
\begin{equation}
    \text{CC}_{\text{Linear}} (f_{\text{in}}, f_{\text{out}}, |\mathcal{V}|)=  2 \cdot f_{\text{in}} \cdot f_{\text{out}} \cdot |\mathcal{V}|
\end{equation}
Suppose the feature vector to the Aggregate-Linear pair (An Aggregate layer followed by a Linear layer) has length $f_{1}$, the output feature vector has length $f_{2}$. The computation complexity of this Aggregate-Linear pair is:
\begin{equation}
  \text{CC}_{\text{Aggregate-Linear}} = 2 \cdot f_{1} \cdot \mathcal{|E|} +2 \cdot f_{1} \cdot f_{2} \cdot \mathcal{|V|} \label{eq:Aggregate-Linear-execution-order}
\end{equation}
If the Aggregate layer and the Linear layer is exchangeable, the computation complexity after the exchange is:
\begin{equation}
    \text{CC}_{\text{Linear-Aggregate}} =2 \cdot f_{1} \cdot f_{2} \cdot \mathcal{|V|} + 2 \cdot f_{2} \cdot \mathcal{|E|} \label{eq:Linear-Aggregate-execution-order}
\end{equation}
\begin{theorem}
\label{theorem:computation-complexity}
Based on Equation (\ref{eq:Aggregate-Linear-execution-order}) and (\ref{eq:Linear-Aggregate-execution-order}),  if $f_{1} >f_{2}$,  Linear-Aggregate execution order has lower complexity. If $f_{2} >f_{1}$ Aggregate-Linear execution order has lower complexity. if $f_{1} = f_{2}$, Aggregate-Linear execution order and Linear-Aggregate execution order have the same computation complexity. 
\end{theorem} 
Based on Theorem \ref{theorem:computation-order} and Theorem \ref{theorem:computation-complexity}, we propose the computation order optimization as shown in Algorithm \ref{algo:computation-order-optimization} (Section \ref{subsec:Computation-Order-Optimization}).
 We iteratively apply Algorithm \ref{algo:computation-order-optimization} until no layers can be exchanged.

 \begin{algorithm}
 \caption{Computation Order Optimizaiton}
 \begin{small}
 \begin{algorithmic}[1]
 \renewcommand{\algorithmicrequire}{\textbf{Input:}}
\renewcommand{\algorithmicensure}{\textbf{Output:}}
 \Require IR of input GNN model, $L$: number of layers in IR
 \Ensure Optimized IR
\For{$l\leftarrow 1$ to $L$ }
    \State{{\color{blue}\# Sequentially check the following conditions}}
    \State \text{Check}: \text{If layer $l$ has only one child layer: layer $m$}
    \State \text{Check}: \text{If layer $m$ has only one parent layer: layer $l$}
    \State \text{Check}: \text{If layer $l$, $m$ is a \{Aggregate, Linear\} pair}
    \State \text{Check}: \text{If the operator of the  Aggregate layer is linear}
    \State \text{Check}: \text{If exchanging layer $l$, $m$ reduces computation}
    \State \hspace*{3.0em}  complexity
    \State{{\color{blue}\# Perform conditional computation order exchange}}
    \If{all the above conditions are met}
        \State Exchange layer $l$ and layer $m$ in IR 
    \EndIf
\EndFor
  \end{algorithmic} 
  \end{small}
 \label{algo:computation-order-optimization}
 \end{algorithm}

\subsection{Layer Fusion}
\label{subsec:layer-fusion}

After computation order optimization, the compiler performs layer fusion consisting of two types: Activation Fusion and BatchNorm Fusion.

\vspace{0.1cm}
\noindent \textbf{Activation Fusion}: An Activation layer can be merged into its adjacent layer, including Aggregate layer, Linear layer, Vector-Inner layer, or Vector-Add layer. Through Activation Fusion, no independent Activation layer is required, which eliminates the external memory traffic between this Activation layer and its adjacent layer.

\vspace{0.1cm}
\noindent \textbf{BatchNorm Fusion}: For inference, the coefficients ($\mu, \sigma, \epsilon, \gamma, \beta$) in the element-wise batch normalization operation are fixed:
$
    y = \frac{x - \mu}{\sqrt{\sigma^{2} + \epsilon}}\cdot \gamma  + \beta
$.
Moreover, the batch normalization operation is a linear operator. Therefore, the BatchNorm layer can be merged with the adjacent Linear layer. The Linear layer incorporates the batch normalization operation into its weights and bias. After BatchNorm Fusion, the BatchNorm layer is eliminated, which reduces total computation complexity and external memory traffic. After layer fusion, the number of computation layers and the computation order of the layers are determined.

\subsection{Data Partitioning}
\label{subsec:data-partitioing}
\begin{figure}[h]
     \centering
     \includegraphics[width=8cm]{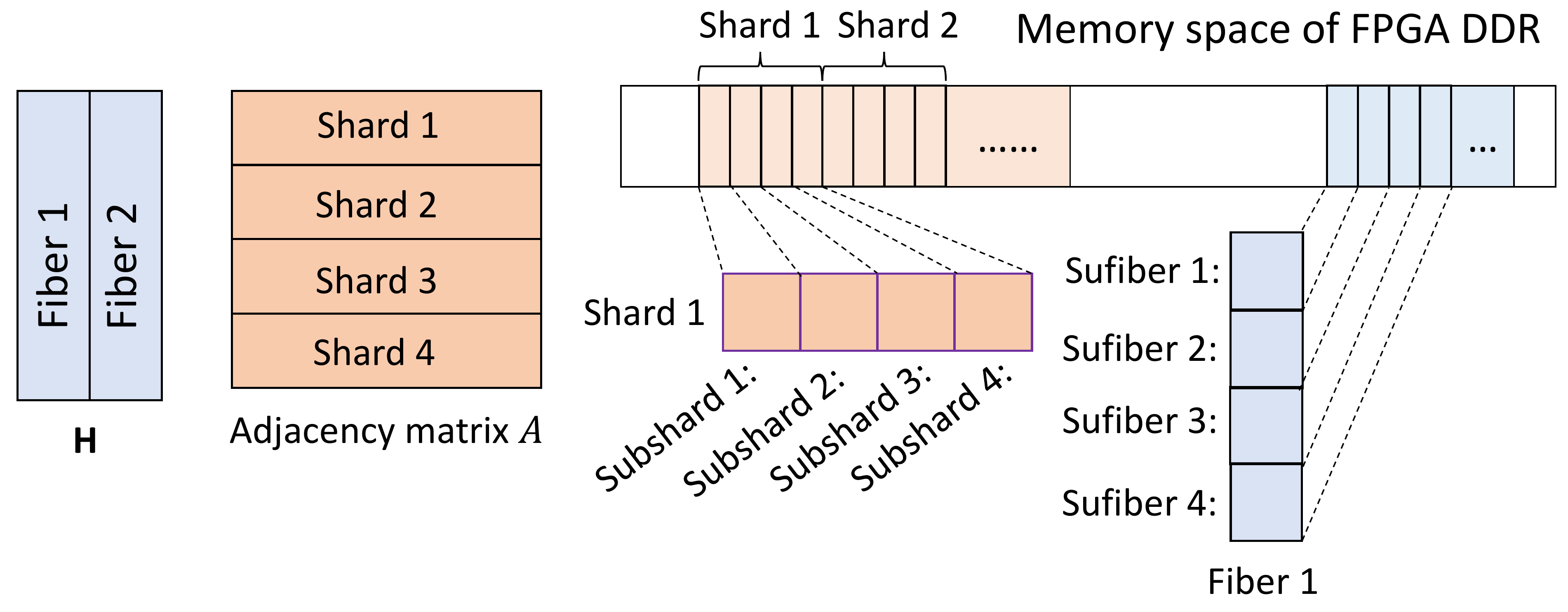}
     \caption{Data partitioning and memory mapping}
     \label{fig:partition-memory-mapping}
\end{figure}

In real-world applications, input graphs can be very large. The compiler performs data partitioning for each layer, starting from the first layer to the last layer. We propose the Fiber-Shard data partitioning (Figure \ref{fig:partition-memory-mapping}) to fit the available on-chip memory. In each layer, the graph has an adjacency matrix $\bm{A}\in \mathbb{R}^{|\mathcal{V}|\times |\mathcal{V}|}$ and a feature matrix $\bm{H}\in \mathbb{R}^{|\mathcal{V}|\times f}$ that need to be partitioned. $\bm{A}$ contains all the edges and is partitioned to shards along the row dimension. Each shard has $N_{\text{1}}$ rows and is partitioned into subshards, with each subshard having $N_{\text{1}}$ columns. The edges in a subshard are stored sequentially in DDR memory, and the subshards in a shard are stored in the contiguous region of DDR memory, as shown in Figure \ref{fig:partition-memory-mapping}. The Feature matrix $\bm{H}$ is partitioned into fibers along column dimension, and each fiber is assigned $N_{2}$ columns. Each fiber is further partitioned into subfibers, and each subfiber has  $N_{1}$ rows. For simplicity, $\bm{A}(i,j)$ denotes the subshard $j$ of shard $i$. $\bm{H}(i,j)$ denotes the subfiber $j$ of fiber $i$.  The same partitioning configuration ($N_{1}$, $N_{2}$) is applied to each layer. The proposed partitioning strategy enables the proposed partition-centric execution scheme (Algorithm \ref{algo:PCFA}, \ref{algo:Vector-Inn}, \ref{algo:Vector-Add}), which further ensures that the outputs of a layer maintain the same partitioning configuration ($N_{1}$, $N_{2}$) as the input. Therefore, the outputs of a layer can be directly used as the input for the next layer since each layer has the same partitioning configuration. Therefore, no data re-partitioning is required between layers. 


\noindent \textbf{Partition-Centric execution scheme}: Based on the Fiber-Shard data partitioning, we propose the partition-centric execution scheme that the execution of a layer is decomposed into a sequence of operations that operate on the data tiles (subshard or subfiber). For example, the execution of an Aggregate layer is described in Algorithm \ref{algo:PCFA}. 
The proposed partition-centric execution scheme leads to reduced memory traffic and random memory access. For the detailed theoretical and empirical analysis of executing the Aggregate layer,  please see our previous work \cite{zhang2021boostgcn}. The proposed partition-centric execution scheme has the following benefits: (1) it enables our block-based kernel mapping (Section \ref{subsec:Kernel-Mapping-task-scheduling}) where each Tiling Block can be executed by a PE independently, and there is no data dependency among Tiling Blocks within a layer, and (2) it enables the unified dynamical task scheduling for each computation layer (Section \ref{subsec:Kernel-Mapping-task-scheduling}).

\begin{algorithm}
 \caption{Partition-Centric execution scheme of an Aggregate Layer}
 \begin{small}
 \begin{algorithmic}[1]
 \renewcommand{\algorithmicrequire}{\textbf{Input:}}
\renewcommand{\algorithmicensure}{\textbf{Output:}}
 \Require $\bm{A}$, $\bm{H}_{\text{in}}$, partitioning configuration ($N_{1}, N_{2}$)
 \Ensure $\bm{H}_{\text{out}}$
\State \tikzmark{start2} \emph{Execution of an Aggregate layer}
\For{$i\leftarrow 1$ to $\frac{f_{\text{in}}}{N_{2}}$ }
    \For{$j\leftarrow 1$ to $\frac{|V|}{N_{1}}$ }  
        \If{there is an idle PE: $\text{PE}_{p}$}
            \State \tikzmark{start1}  \hspace{-0.2cm}  Assign $\bm{H}_{\text{out}}(i, j)$ to $\text{PE}_{p}$
            \State   Initialize $\bm{H}_{\text{out}}(i, j)$
                \For{$k\leftarrow 1$ to $\frac{|V|}{N_{1}}$ }
                    \State load  $\bm{A}(j, k)$ to Edge Buffer
                    \State load $\bm{H}_{\text{in}}(k, i)$ to Feature Buffer
                    \State  $\bm{H}_{\text{out}}(i, j) \leftarrow \text{SpDMM}(\bm{A}(j, k), \bm{H}_{\text{in}}(k, i))$
                \EndFor
            \State  Apply activation if required 
            \State Store $\bm{H}_{\text{out}}(i, j)$  
            \tikzmark{end1} \tikzmark{end2}
        \EndIf 
    \EndFor
\EndFor 
 \end{algorithmic} 
 \end{small}
 \label{algo:PCFA} 
 \end{algorithm}
\Textbox[4.4cm]{start1}{end1}{Tiling Block}
\Textbox[4.3cm]{start2}{end2}{Layer Block}

\vspace{0.1cm}
\noindent \textbf{Data partitioning of a Linear layer}: A Linear layer performs  matrix multiplication of input feature matrix $\bm{H}_{\text{in}}\in \mathbb{R}^{|\mathcal{V}|\times f_{in}}$ and weight matrix $\bm{W}\in \mathbb{R}^{f_{\text{in}}\times f_{\text{out}}}$. Output feature matrix is $\bm{H}_{\text{out}} = \bm{H}_{\text{in}}\bm{W}$. For the Linear layer, we perform the standard block matrix multiplication. For the Linear layer, the data partitioning keeps the same partitioning configuration as described on Section \ref{subsec:data-partitioing} for the input feature matrix $\bm{H}_{\text{in}}$ and output feature matrix $\bm{H}_{\text{out}}$. The basic computation kernel of a Linear layer is the GEMM.

\vspace{0.1cm}
\noindent \textbf{Data partitioning of a Vector-Inn layer}: A Vector-Inn layer is to sample the results using adjacent matrix $\bm{A}_{\text{in}}$ from $ (\bm{H_{\text{in}}}\bm{H}_{\text{in}}^{T})$, which is denoted as $\bm{A}_{\text{in}} \odot (\bm{H_{\text{in}}}\bm{H}_{\text{in}}^{T})$. The output $\bm{A}_{\text{out}}$ is the combination of $\bm{A}_{\text{in}}$ and the weight value of each non-zero position in $\bm{A}_{\text{in}}$. The Vector-Inn layer exploits the same partitioning strategy (See Section \ref{subsec:data-partitioing}) as the Aggregate Layer. The execution scheme of a Vector-Inn layer is shown in Algorithm \ref{algo:Vector-Inn}.

\begin{algorithm}
 \caption{Partition-Centric execution scheme of a Vector-Inn Layer}
 \begin{algorithmic}[1]
 \renewcommand{\algorithmicrequire}{\textbf{Input:}}
\renewcommand{\algorithmicensure}{\textbf{Output:}}
 \Require $\bm{A}_{\text{in}}$, $\bm{H}_{\text{in}}$
 \Ensure $\bm{A}_{\text{out}}$
\State \tikzmark{start2} \emph{Execution of a Vector-Inn layer}
\For{ $i\leftarrow 1$ to  $\frac{|V|}{N_{1}}$ } 
    \For{$j\leftarrow 1$ to $\frac{|V|}{N_{1}}$ } 
        \If{there is an idle PE: $\text{PE}_{p}$}
            \State Assign $\bm{A}_{\text{out}}(i, j)$  to $\text{PE}_{p}$
            \State  \tikzmark{start1} \hspace{-0.2cm}  Initialize $\bm{A}_{\text{out}}(i, j)$ 
            \State load  $\bm{A}_{\text{in}}(i, j)$ to Edge Buffer
                \For{$k\leftarrow 1$ to $\frac{f_{\text{in}}}{N_{2}}$}
                    \State load  $\bm{H}_{\text{in}}(i, k)$ to Feature Buffer
                    \State load $\bm{H}_{\text{in}}(j, k)$ to Feature Buffer
                    \State  \begin{small}
                    $\bm{Z}\leftarrow \text{SDDMM}(\bm{A}_{\text{in}}(i, j), \bm{H}_{\text{in}}(i, k), \bm{H}_{\text{in}}(j, k))$ \end{small}
                    \State $\bm{A}_{\text{out}}(i, j)\leftarrow \text{Apply}(Z)$
                    \State  Apply activation if required
                \EndFor
            \State Store $\bm{A}_{\text{out}}(i, j)$  \tikzmark{end1} \tikzmark{end2}
        \EndIf
    \EndFor
\EndFor 
  \end{algorithmic} 
 \label{algo:Vector-Inn}
 \end{algorithm}
\Textbox[4.4cm]{start1}{end1}{Tiling Block}
\Textbox[4.3cm]{start2}{end2}{Layer Block}

\vspace{0.1cm}
\noindent \textbf{Data partitioning of a Vector-Add layer}: The inputs to a Vector-Add layer are two input feature matrices of the same size -- $\bm{H}_{\text{in}}^{l_{1}}$ and  $\bm{H}_{\text{in}}^{l_{2}}$. The output feature matrix $\bm{H}_{\text{out}}$ is the addition of two matrices: $\bm{H}_{\text{out}} = \bm{H}_{\text{in}}^{l_{1}} + \bm{H}_{\text{in}}^{l_{2}}$. The execution of the Vector-Add layer is shown in Algorithm \ref{algo:Vector-Add}.

\begin{algorithm}
 \caption{Partition-Centric  execution scheme of an Vector-Add Layer}
 \begin{algorithmic}[1]
 \renewcommand{\algorithmicrequire}{\textbf{Input:}}
\renewcommand{\algorithmicensure}{\textbf{Output:}}
 \Require $\bm{H}_{\text{in}}^{l_{1}}$, $\bm{H}_{\text{in}}^{l_{2}}$
 \Ensure $\bm{H}_{\text{out}}$
 \State \tikzmark{start4} \emph{Execution of a Vector-Add layer}
\For{$i\leftarrow 1$ to $\frac{f_{\text{in}}}{N_{2}}$ } 
    \For{$j\leftarrow 1$ to $\frac{|V|}{N_{1}}$ } 
        \If{there is an idle PE: $\text{PE}_{p}$}
            \State Assign $\bm{H}_{\text{out}}(i, j)$ to $\text{PE}_{p}$
            \State \tikzmark{start3}  \hspace{-0.2cm}   load  $\bm{H}_{\text{in}}^{l_{1}}(i, j)$ to Feature Buffer
            \State load  $\bm{H}_{\text{in}}^{l_{2}}(i, j)$ to Feature Buffer

            \State $\bm{H}_{\text{out}}(i, j) \leftarrow \text{VecAdd}(\bm{H}_{\text{in}}^{l_{1}}(i, j), \bm{H}_{\text{in}}^{l_{2}}(i, j))$
            \State Store $\bm{H}_{\text{out}}(i, j)$  \tikzmark{end3} \tikzmark{end4}
        \EndIf
    \EndFor
\EndFor
\end{algorithmic} 
\label{algo:Vector-Add}
\end{algorithm}
 \Textbox[4.4cm]{start3}{end3}{Tiling Block}
\Textbox[4.3cm]{start4}{end4}{Layer Block}


\subsection{Kernel Mapping and Task Scheduling}
\label{subsec:Kernel-Mapping-task-scheduling}

\noindent \textbf{Kernel Mapping}: Through data partitioning, each layer in the IR is expressed as nested loops (e.g., Algorithm \ref{algo:PCFA}) according to the proposed partition-centric execution scheme. The compiler maps each layer to a sequence of high-level instructions. The kernel mapping is performed hierarchically. Each layer is mapped to a block of instructions called \textbf{Layer Block} (e.g., Algorithm \ref{algo:PCFA}). Each Layer Block contains a Control and Scheduling Instruction (CSI) and a set of \textbf{Tiling Blocks}. The Tiling Blocks are generated by unfolding the outer nested loops of a Layer Block. For example, for an Aggregate layer, the generated CSI contains the information of Line 2-3 in Algorithm \ref{algo:PCFA}, and $\frac{f_{\text{in}}}{N_{2}}\times \frac{|V|}{N_{1}}$ Tiling Blocks are generated by unfolding the outer loops. A Tiling Block has an inseparable sequence of high-level instructions that will be executed by a PE.

\begin{algorithm}
 \caption{Task Scheduling}
  \begin{small}
 \begin{algorithmic}[1ht]
 \renewcommand{\algorithmicrequire}{\textbf{Input:}}
\renewcommand{\algorithmicensure}{\textbf{Output:}}
 \Require $\bm{A}$ and $\bm{H}_{\text{in}}$ of input graph; weight matrices; $L$: number of Layer Blocks.
 \Ensure output embedding of each vertex
\For{$l \leftarrow 1$ to $L$ }
    \State Load CSI of Layer Block $l$ to Scheduler
    \For{each Tiling block in Layer Block $l$ \textbf{parallel}}
        \If{there is an idle PE: $\text{PE}_{p}$}
            \State Assign this Tiling Block to $\text{PE}_{p}$
            \State $\text{PE}_{p}$ Executes this Tiling Block
        \EndIf
    \EndFor 
    \State Wait until all the Tiling Blocks are executed
\EndFor 
  \end{algorithmic} 
  \end{small}
 \label{algo:task-scheduling} 
 \end{algorithm}

\vspace{0.1cm}
\noindent \textbf{Task Scheduling}: As shown in Algorithm  \ref{algo:task-scheduling}, GraphAGILE executes the GNN inference layer by layer. For each Layer Block, the Scheduler loads the heading Control and Scheduling Instruction (CSI). Then, the Scheduler assigns the Tiling Blocks to the idle PEs, forming a \emph{dynamic load balancing strategy}. Each PE maintains a 1-bit output port to indicate its current status (Idle/Busy). When all the Tiling Blocks within a layer are completely finished, GraphAGILE starts to execute the next layer. Within each Tiling Block, the computation instructions and memory read/write instructions are interleaved. Therefore, we exploit the double buffering technique to overlap the computation and data communication. Specifically, Instruction Decoder \& Control Signal Generator needs to issue new memory read instructions when the old computation instruction is not finished, which may incur  Write after Read (WAR) data hazard. Therefore, each buffer in a PE maintains a hardware mutex implemented as a one-bit register. After a memory read instruction loads data to a buffer, it locks the mutex of this buffer. After the computation instruction finishes using the data from this buffer, the mutex is unlocked. When a memory read instruction is stalled by a lock, the Instruction Decoder \& Control Signal Generator will stop issuing new instructions. Locking/unlocking the mutex is annotated in the high-level instructions by the compiler. Such annotation is through scanning the data dependency among high-level instructions within each Tiling Block, which has negligible complexity.
After kernel mapping and mutex annotation, the compiler generates the executable file.

\vspace{0.4cm}
\vspace{0.4cm}
\section{Implementation Details}
\label{Sec:Experimental-Evaluation}

{We conduct comprehensive experiments to evaluate the performance of GraphAGILE. Section \ref{Sec:Experimental-Evaluation} introduces the implementation details and experimental settings, while Section \ref{sec:Evaluation-Results} presents the detailed experimental results, including the execution time (e.g., end-to-end latency, latency of compilation, and latency of hardware execution) (Section \ref{subsec:execution-time}), the impact of compiler optimizations (Section \ref{subsec:Impact-of-the-Optimizations}), cross-platform comparison (Section \ref{subsec:cross-platform-comparsion}), and the comparison with state-of-the-art accelerators (Section \ref{subsec:Comparison-State-of-The-Art}).}


\begin{figure}[h]
     \centering
     \includegraphics[width=7.5cm]{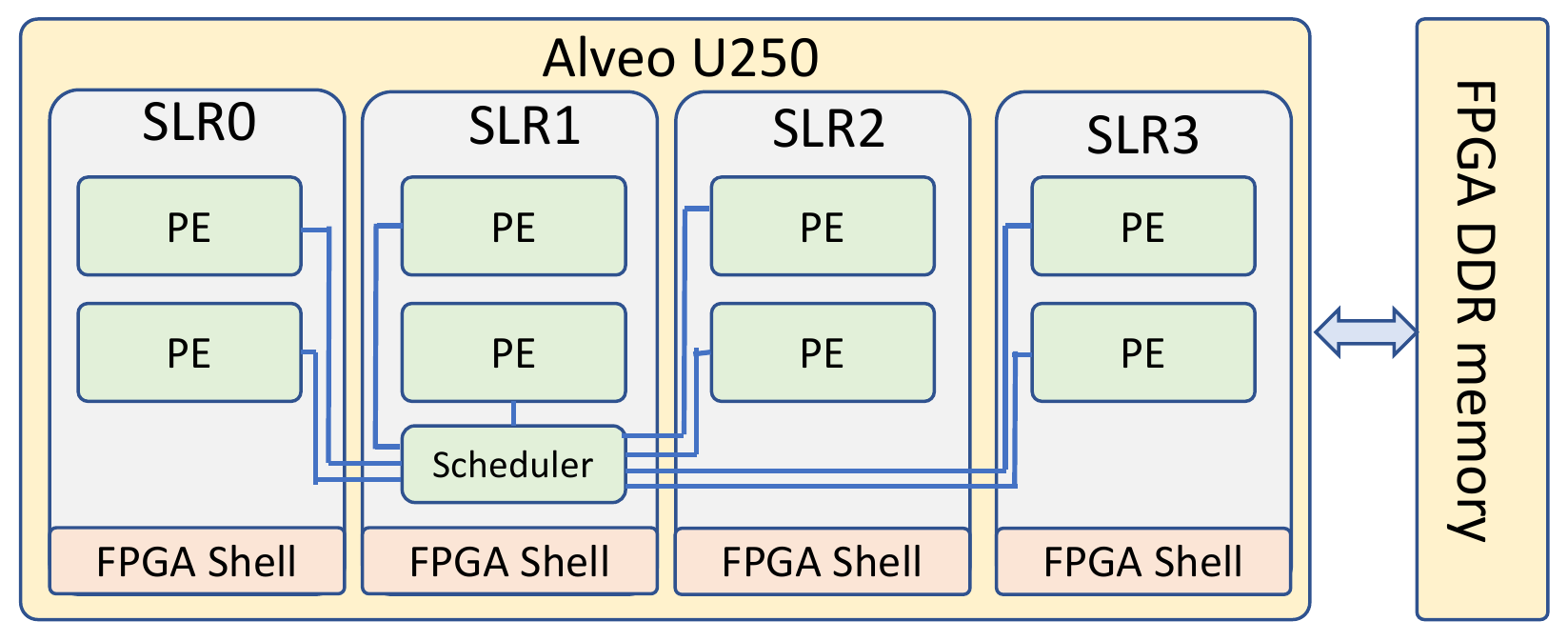}
     \caption{The mapping of GraphAGILE on Alveo U250}
     \label{fig:platform-mapping}
\end{figure}

\begin{figure*}[h]
     \centering
     \includegraphics[width=18cm]{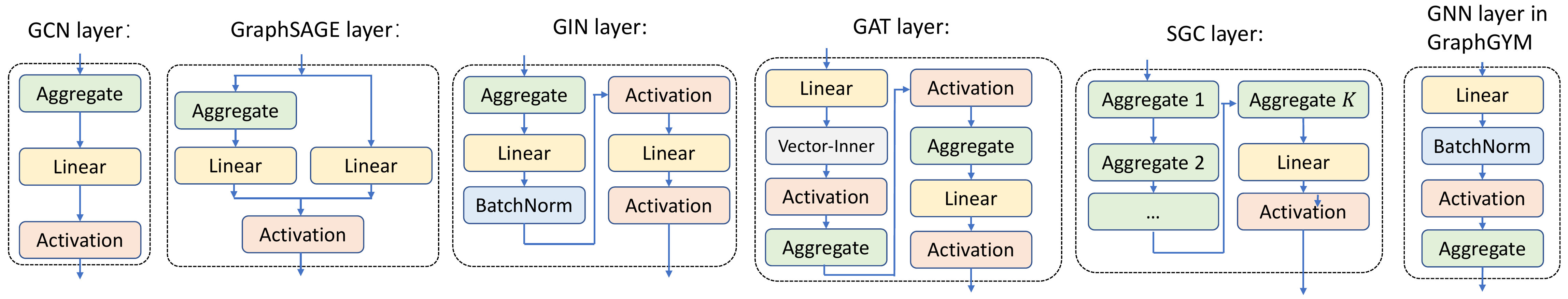}
     \caption{Intermediate Representations of state-of-the-art GNN layers}
     \label{fig:IR-state-of-the-art}
\end{figure*}

We implement the hardware design on a state-of-the-art FPGA platform, Xilinx Alveo U250, consisting of four Super Logic Regions (SLRs). The FPGA DDR memory has four channels with 77 GB/s memory bandwidth. On U250, we implement 8 PEs where each SLR contains 2 PEs of $p_{\text{sys}}=16$.  
We develop GraphAGILE using {Verilog HDL}. We synthesize the design and perform Place\&Route using Xilinx Vivado 2021.1 to obtain the frequency and resource utilization report.
GraphAGILE on Alveo U250 consumes  $778$K LUTs (45\%), $10240$ DSPs (83\%), $1853$ BRAMs (69\%) and $1050$ URAMs (82\%). GraphAGILE runs at 300 MHz.
Then, we build a cycle-accurate simulator for the hardware accelerator to evaluate its performance. We use Ramulator \cite{kim2015ramulator} to simulate the performance of FPGA DDR memory. 
We develop the compiler using Python. At runtime, the compiler reads the user-defined GNN models (defined using Pytorch Geometric library (PyG) \cite{paszke2019pytorch}) and input graphs. Then, the compiler generates the binary file for the hardware accelerator and performs preprocessing for the input graph. After that, the binary file, GNN model weights, and propocessed input graph are sent to the FPGA DDR memory through PCIe. For performance simulation, we set the PCIe bandwidth to be $31.5$ GB/s which is the same as the baseline CPU-GPU platform for a fair comparison. {The Alveo U250 board has four DDR memories, each connected to an SLR, and each DDR memory has capacity of 16 GB. The DDR memory on the Alveo U250 board is sufficient to store the input graphs used in our experiments (Table \ref{tab:datasets-statistics}). For example, for the largest graph used in our experiments, Amazon-Products, has a total size of 7.2 GB, including the vertex features and the edges.}

\begin{figure}[h]
     \centering
     \includegraphics[width=7cm]{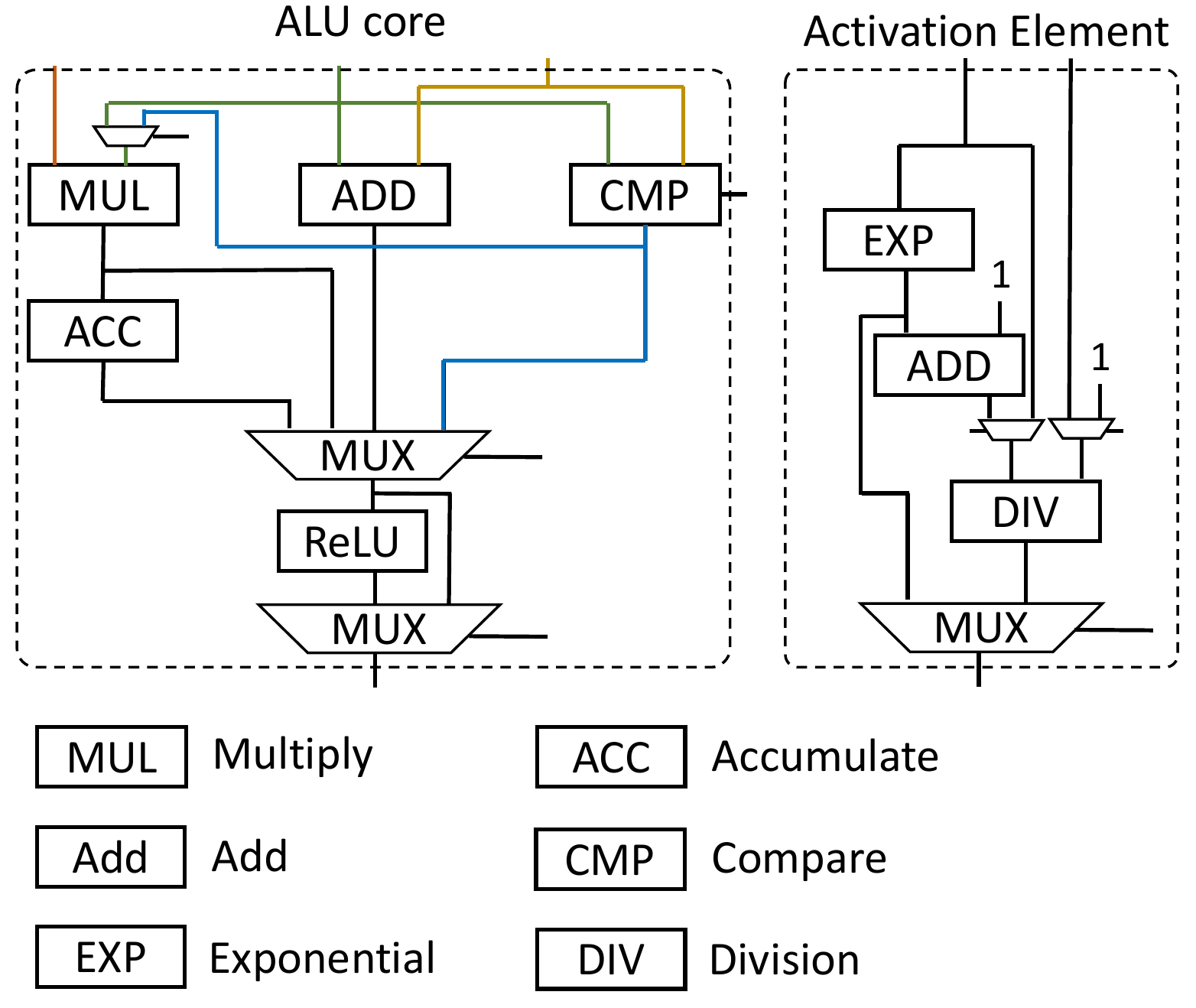}
     \caption{The structure of the ALU in ACK and the structure of an Activation Element in Activation Unit}
     \label{fig:structure-ALU}
\end{figure}

\noindent \textbf{Arithmetic Logic Unit (ALU)}: The proposed ALU can support multiplication, addition, multiply-add operation, comparison (Mux, Min), ReLU activation, PReLU activation. Each PE also has an Activation Unit and the Activation Unit has 16 Activation Elements (See Figure \ref{fig:structure-ALU}) that work in parallel. The Activation Element supports exponential function $\verb|exp(x)|$, sigmoid function $\verb|1/(1+exp(x))|$, division. 


\begin{figure}[h]
     \centering
     \includegraphics[width=8.5cm]{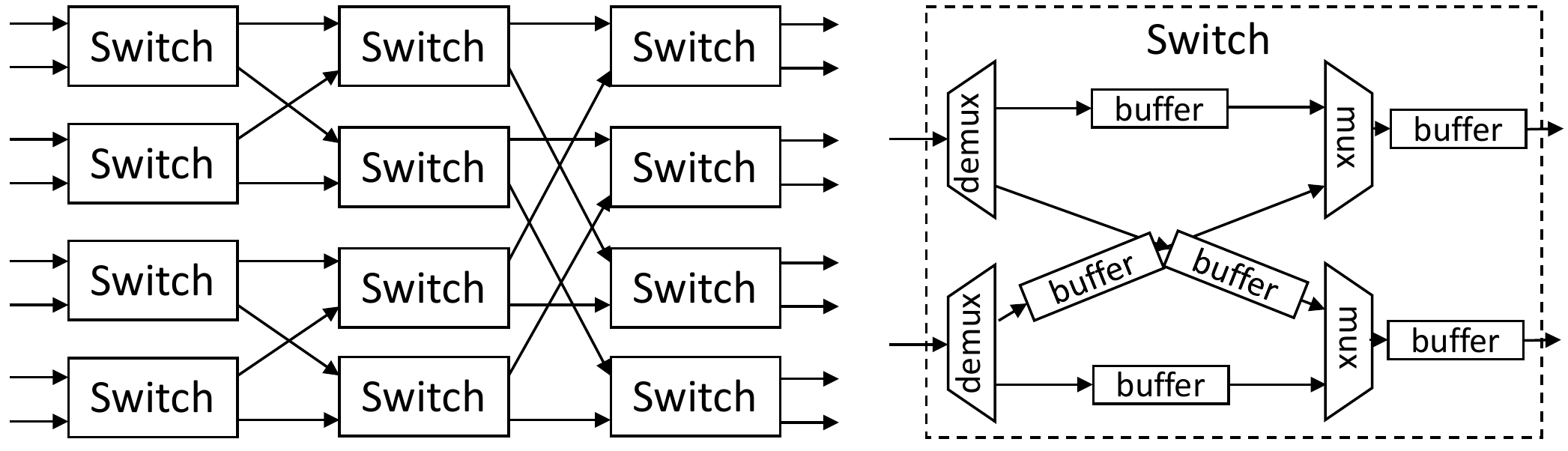}
     \caption{The structure of ISN and DSN ($p_{\text{sys}} = 8$)}
     \label{fig:Bufferfly}
\end{figure}

\vspace{0.1cm}
\noindent \textbf{Routing Network}: The proposed Index Shuffle Network (ISN) and Data Shuffle Network (DSN) are implemented using Bufferfly Network that is proposed in \cite{choi2021hbm}. The structure of the Bufferfly Network is depicted in Figure \ref{fig:Bufferfly}. The benefits of using this Bufferfly Network are: (1) the Bufferfly Network is hardware efficient that only consumes a small number of ALUs, (2) there are intermediate buffers in the Switches that can buffer data when there is network congestion. As {shown} in \cite{choi2021hbm}, such intermediate buffers lead to high-throughput data routing.

\begin{figure}[h]
  \centering
    \includegraphics[width=4cm]{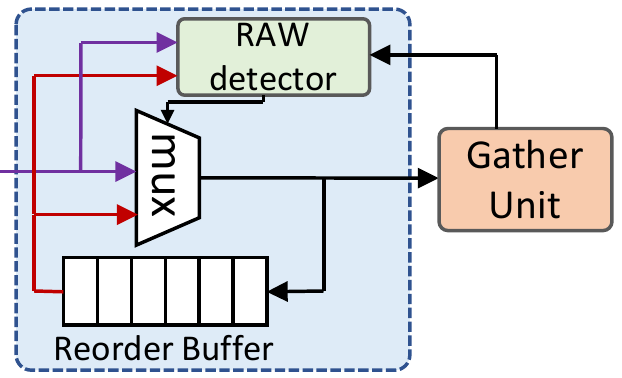}
    \captionof{figure}{RAW Unit}
     \label{fig:RAWUnit-Arch}
\end{figure}

\vspace{0.1cm}
\noindent \textbf{RAW Unit}:  In SpDMM mode, read-after-write (RAW) data hazard may occur when accumulators in the Gather Unit read the old feature vertex vector from the Feature/Result Buffer.
To resolve the RAW data hazard, we implement a RAW Unit before Gather Unit as shown in Figure \ref{fig:RAWUnit-Arch}. In the RAW Unit, there is a RAW detector to detect the RAW data hazard and a small Reorder Buffer (implemented as a FIFO) to cache the input data when RAW is detected. The data in the Reorder Buffer will be sent to the Gather Unit when there is no RAW data hazard.

\vspace{0.1cm}
\noindent \textbf{System Details of Alveo U250}:  {Figure \ref{fig:structure-ALU} depicts} the structure of an ALU in the ACK.   Each PE has an Index Shuffle Network and a Data Shuffle Network. The Index Shuffle Network has 16 input ports and 16 output ports. Each port has 96 bits since an edge is  96-bit (32-bit source index, 32-bit destination index, 32-bit edge weight). The Data shuffle network also has 16 input ports and 16 output ports. Each port has $(16+3)\times32$ bits where $16 \times 32$ bits are used for vector features and $3 \times 32$ bits are used for edges. Both Index Shuffle Network and Data Shuffle Network are implemented using a butterfly network. Each PE has an Edge Buffer of size 2MB, a Feature Buffer of size 3MB, and a Weight Buffer of size 1MB. We exploit the double buffering technique for Edge Buffer and Weight Buffer and the triple buffering technique for Feature Buffer. Such triple/double buffering enables the overlapping of computation and data communication. The dimension of a Weight Buffer is $(N_{W} = 16384)\times (P_{\text{sys}}=16)$, the dimension of an Edge Buffer is $(N_{E}=65K)\times 3$, the dimension of a Feature Buffer is $(N_{F1}=16384)\times (N_{F2}=16)$.


\vspace{0.1cm}
\noindent \textbf{Resource utilization}:  Table \ref{tab:resource-utilizations} shows the resource utilization of various FPGA accelerators in the experiments.

\begin{table}[!ht]
\centering
\caption{Specifications of FPGA accelerators in the experiments }
\vspace{-0.2cm}
\begin{threeparttable}
\begin{adjustbox}{max width=0.48\textwidth}
\begin{tabular}{c|ccc}
 \toprule
\textbf{Platforms} &  AWB-GCN \cite{geng2020awb} &  BoostGCN \cite{zhang2021boostgcn} & GraphAGILE  \\ 
\midrule \midrule 
Platform  &  Stratix 10 SX &  Stratix 10 GX  & Alveo U250  \\
 {Platform Technology}  &  Intel 14 nm & Intel 14 nm &  TSMC 16 nm\\ 
{Frequency} & 330 MHz  & 250 MHz & 300 MHz \\  
LUTs/ALMs & 200K-300K ALMs & 294K ALMs & 778K LUTs \\
DSPs & 4096 (Intel DSP) & 3840 (Intel DSP) & 10240 (Xilinx DSP) \\
{Peak Performance}& 1351 GFLOPS & 640  GFLOPS & 614 GFLOPS  \\ 
{On-chip Memory}& 22MB & 32 MB & 45 MB   \\
{Memory Bandwidth}& 57.3 GB/s & 77 GB/s  & 77 GB/s \\ \bottomrule
\end{tabular}
\end{adjustbox}
\end{threeparttable}
\label{tab:resource-utilizations}
\end{table}
\section{Evaluation Results} 
\label{sec:Evaluation-Results}

\begin{table}[ht]
\centering
\caption{Dataset Statistics}
\begin{adjustbox}{max width=0.48\textwidth}
\begin{tabular}{ccccc}
\toprule
\textbf{Dataset}   & \textbf{Vertices} & \textbf{Edges} & \textbf{Features} & \textbf{Classes}\\
\midrule
\midrule
Citeseer (CI) \cite{kipf2016semi} & 3327 & 4732& 3703& 6 \\
Cora (CO) \cite{kipf2016semi} & 2708 &  5429& 1433 & 7  \\
Pubmed (PU) \cite{kipf2016semi} & 19717 &  44338 & 500 & 3  \\
Flickr (FL) \cite{zeng2019graphsaint} & 89,250 &    899,756 & 500 & 7 \\
Reddit (RE) \cite{hamilton2017inductive} & 232,965  & 116,069,919 & 602 & 41\\
Yelp (YE) \cite{zeng2019graphsaint} & 716,847  & 6,977,410 & 300 & 100      \\
Amazon-Products (AP) \cite{hu2020open} &  1,569,960 &    264,339,468 & 200 & 107 \\
\bottomrule
\end{tabular}
\end{adjustbox}
\label{tab:datasets-statistics}
\end{table}

\begin{table}[ht]
\centering
\caption{Evaluated GNN models in the experiments}
\begin{adjustbox}{max width=0.48\textwidth}
\begin{tabular}{cccccc}
\toprule
\textbf{Notation} & \textbf{Layer  Type} & \textbf{\# of layers} & \textbf{Hidden Dimension} & \textbf{Ref.} \\ \midrule \midrule
\textbf{b1} & GCN layer & 2 & 16& \cite{kipf2016semi,geng2020awb, yan2020hygcn} \\ \midrule
\textbf{b2} & GCN layer & 2 & 128& \cite{yan2020hygcn,geng2020awb} \\ \midrule
\textbf{b3} & GraphSAGE layer & 2 & 128& \cite{hamilton2017inductive,zeng2019graphsaint}\\ \midrule
\textbf{b4} & GraphSAGE layer & 2 & 256 & \cite{hamilton2017inductive,zeng2019graphsaint}\\ \midrule
\textbf{b5} & GIN layer & 5 & 128-128-128-128 & \cite{xu2018powerful}\\  \midrule
\textbf{b6} & GAT layer & 2 & 64 & \cite{velivckovic2017graph}\\ \midrule
\textbf{b7} & SGC layer & 1 (k=2) & N/A  & \cite{wu2019simplifying}\\  \midrule
\textbf{b8} & GraphGym layer & \begin{tabular}[|c|]{@{}c@{}} 1 preprocessing layer\\  3 GNN layer \\  1 postprocessing layer \end{tabular} & 256  & \cite{you2020design} \\ 
\bottomrule
\end{tabular}
\end{adjustbox}
\label{tab:evaluated-GNN-model}
\end{table}

\begin{table*}[!ht]
\centering
\caption{Specifications of platforms }
\begin{threeparttable}
\begin{adjustbox}{max width=0.95\textwidth}
\begin{tabular}{c|cccccc}
 \toprule
\textbf{Platforms} & CPU & GPU  & HyGCN \cite{yan2020hygcn} & AWB-GCN \cite{geng2020awb} &  BoostGCN \cite{zhang2021boostgcn} & GraphAGILE  \\ 
\midrule \midrule 
Platform  & AMD Ryzen 3990x & Nvidia RTX3090 & ASIC &  Stratix 10 SX &  Stratix 10 GX  & Alveo U250  \\
 {Platform Technology}  & TSMC 7 nm   & TSMC 7 nm & TSMC 12 nm & Intel 14 nm & Intel 14 nm &  TSMC 16 nm\\ 
{Frequency} & 2.90 GHz  & 1.7 GHz & 1 GHz  & 330 MHz  & 250 MHz & 300 MHz \\ 
{Peak Performance}& 3.7 TFLOPS & 36 TFLOPS &  4608 GFLOPS& 1351 GFLOPS & 640  GFLOPS & 614 GFLOPS  \\ 
{On-chip Memory}& 256 MB L3 cache & 6 MB L2 cache & 35.8 MB&  22MB & 32 MB & 45 MB   \\
{Memory Bandwidth}& 107 GB/s & 936.2 GB/s  & 256 GB/s & 57.3 GB/s & 77 GB/s  & 77 GB/s \\ \bottomrule
\end{tabular}
\end{adjustbox}
\end{threeparttable}
\label{tab:platform-specifications}
\end{table*}

Figure \ref{fig:IR-state-of-the-art} shows the IR of various types of widely-used GNN layers evaluated in our experiments.

\vspace{0.1cm}
\noindent \textbf{Baselines}: As shown in Table \ref{tab:platform-specifications}, we compare our design with state-of-the-art baselines: CPU-only platform (AMD Ryzen 3990x), CPU-GPU (AMD Ryzen 3990x + Nvidia RTX3090), HyGCN \cite{yan2020hygcn}, BoostGCN \cite{zhang2021boostgcn}, AWB-GCN \cite{geng2020awb}.

\vspace{0.1cm}
\noindent \textbf{Benchmarks}: We use eight GNN models in Table \ref{tab:evaluated-GNN-model} and seven graph datasets in Table \ref{tab:datasets-statistics} {as benchmarks}.

\vspace{0.1cm}
\noindent \textbf{Performance Metric}: We evaluate the performance by:
\begin{itemize}
    \item \emph{End-to-End (E2E) latency $T_{\text{E2E}}$}:  
    The $T_{\text{E2E}}$ of GraphAGILE includes (1) the latency of software compilation $ T_{\text{LoC}}$ on the host processor, (2) the latency of CPU-FPGA data movement  $T_{\text{comm}}$, and (3) the latency of executing GNN inference on the accelerator (Latency of hardware execution  $T_{\text{LoH}}$). The latency of moving data (processed graph, GNN model, binary file) from host platform to FPGA DDR  $T_{\text{comm}}$ is estimated through:
$
    T_{\text{comm}} = \frac{\text{total data volume}}{\text{sustained PCIe bandwidth}}
$. 
Then, the end-to-end latency of GraphAGILE is calculated by:
$
    T_{\text{E2E}} = T_{\text{LoC}} + T_{\text{comm}} + T_{\text{LoH}}
$.

    \item \emph{Latency of compilation (LoC) $T_{\text{LoC}}$}:  The latency of compilation is the overhead of software or hardware compilation. The measured $T_{\text{LoC}}$ of GraphAGILE is the time duration \emph{from} the time the GNN model (defined using PyG API) and the input graph are provided, \emph{to} the time the input graph is processed and the instruction sequence is generated by the compiler. For the design automation frameworks \cite{zhang2021boostgcn, liang2020deepburning}, $T_{\text{LoC}}$ includes hardware meta compilation, hardware synthesis, Place\&Route, and FPGA reconfiguration. 
    \item \emph{Latency of hardware execution (LoH) $T_{\text{LoH}}$}: Latency of hardware execution is the latency of executing the binary code on the hardware accelerator. Before runtime, the GNN model, processed input graph, and binary file are already stored in the FPGA DDR. 
\end{itemize}




\subsection{Execution Time and Size of Binary File}
\label{subsec:execution-time}



\begin{scriptsize}
\begin{table}[ht]
\centering
\caption{End-to-End latency, latency of compilation, latency of hardware execution}
\begin{adjustbox}{max width=0.48\textwidth}
\begin{tabular}{ccccccccc}
\toprule
\multirow{2}{*}{\textbf{Model}} & \multirow{2}{*}{\textbf{\begin{tabular}[|c|]{@{}c@{}}Latency \\ (ms)\end{tabular}}} & \multicolumn{7}{c}{\textbf{Dataset}} \\ \cmidrule{3-9}
& & CI&CO&PU&FL&RE&YE&AP\\ \midrule
\textbf{b1}&\begin{tabular}[|c|]{@{}c@{}}$T_{\text{E2E}}$\\ $T_{\text{LoC}}$  \\  $T_{\text{LoH}}$ 
\end{tabular}
& \begin{tabular}[|c|]{@{}c@{}} 2.129 \\ 0.249 \\ 0.320 \end{tabular}  
& \begin{tabular}[|c|]{@{}c@{}} 0.808 \\ 0.215 \\ 0.103 \end{tabular}  
& \begin{tabular}[|c|]{@{}c@{}} 2.126 \\ 0.574 \\ 0.272 \end{tabular}  
& \begin{tabular}[|c|]{@{}c@{}} 9.97 \\ 2.68 \\ 1.28 \end{tabular}  
& \begin{tabular}[|c|]{@{}c@{}} 128.3 \\ 51.1 \\ 15.6 \end{tabular}  
& \begin{tabular}[|c|]{@{}c@{}} 62.9 \\ 18.8 \\ 11.6 \end{tabular}  
& \begin{tabular}[|c|]{@{}c@{}} 442.0 \\ 263.8 \\ 37.4 \end{tabular}  
\\ \midrule
\textbf{b2}&\begin{tabular}[|c|]{@{}c@{}} $T_{\text{E2E}}$\\$T_{\text{LoC}}$ \\  $T_{\text{LoH}}$ \end{tabular}
& \begin{tabular}[|c|]{@{}c@{}} 4.364\\ 0.254 \\ 2.550 \end{tabular}  
& \begin{tabular}[|c|]{@{}c@{}} 1.535 \\ 0.226 \\ 0.819 \end{tabular}  
& \begin{tabular}[|c|]{@{}c@{}} 4.28 \\ 0.66 \\ 2.34 \end{tabular}  
& \begin{tabular}[|c|]{@{}c@{}} 20.1 \\ 2.6 \\ 11.5 \end{tabular}  
& \begin{tabular}[|c|]{@{}c@{}} 208.5 \\ 49.7 \\ 97.2 \end{tabular}  
& \begin{tabular}[|c|]{@{}c@{}} 155.1 \\ 18.3 \\ 104.3 \end{tabular}  
& \begin{tabular}[|c|]{@{}c@{}} 718.1 \\ 261.4 \\ 315.9 \end{tabular}  
\\ \midrule
\textbf{b3}&\begin{tabular}[|c|]{@{}c@{}}$T_{\text{E2E}}$\\$T_{\text{LoC}}$ \\  $T_{\text{LoH}}$
\end{tabular}
& \begin{tabular}[|c|]{@{}c@{}} 4.355 \\ 0.235 \\ 2.560 \end{tabular}  
& \begin{tabular}[|c|]{@{}c@{}} 1.574  \\ 0.258 \\ 0.826 \end{tabular}  
& \begin{tabular}[|c|]{@{}c@{}} 4.25 \\ 0.59 \\ 2.38 \end{tabular}  
& \begin{tabular}[|c|]{@{}c@{}} 21.19 \\ 2.58 \\ 12.60 \end{tabular}  
& \begin{tabular}[|c|]{@{}c@{}} 212.7 \\ 49.1 \\ 102.0 \end{tabular}  
& \begin{tabular}[|c|]{@{}c@{}} 134.3 \\ 19.2 \\ 82.6 \end{tabular}  
& \begin{tabular}[|c|]{@{}c@{}} 657.4 \\ 272.2 \\ 244.4 \end{tabular}  
\\ \midrule
\textbf{b4}&\begin{tabular}[|c|]{@{}c@{}}$T_{\text{E2E}}$\\ $T_{\text{LoC}}$ \\ $T_{\text{LoH}}$ 
\end{tabular}
& \begin{tabular}[|c|]{@{}c@{}} 6.912 \\ 0.212 \\ 5.140 \end{tabular}  
& \begin{tabular}[|c|]{@{}c@{}} 2.387 \\ 0.237 \\ 1.660 \end{tabular}  
& \begin{tabular}[|c|]{@{}c@{}} 6.919 \\ 0.599 \\ 5.040 \end{tabular}  
& \begin{tabular}[|c|]{@{}c@{}} 33.88 \\ 2.47 \\ 25.40 \end{tabular}  
& \begin{tabular}[|c|]{@{}c@{}} 315.0 \\ 50.1 \\ 203.3 \end{tabular}  
& \begin{tabular}[|c|]{@{}c@{}} 278.2 \\ 21.3 \\ 224.4 \end{tabular}  
& \begin{tabular}[|c|]{@{}c@{}} 905.2 \\ 270.3 \\ 494.1 \end{tabular}  
\\ \midrule
\textbf{b5}&\begin{tabular}[|c|]{@{}c@{}} $T_{\text{E2E}}$ \\ $T_{\text{LoC}}$  \\ $T_{\text{LoH}}$ \end{tabular}
& \begin{tabular}[|c|]{@{}c@{}} 14.99 \\ 0.24 \\ 13.10 \end{tabular}  
& \begin{tabular}[|c|]{@{}c@{}} 9.23 \\ 0.23 \\ 8.51 \end{tabular}  
& \begin{tabular}[|c|]{@{}c@{}} 15.64 \\0.56 \\ 13.80 \end{tabular}  
& \begin{tabular}[|c|]{@{}c@{}} 91.73 \\ 2.52 \\ 83.20 \end{tabular}  
& \begin{tabular}[|c|]{@{}c@{}} 527.6 \\ 50.9 \\ 415.1 \end{tabular}  
& \begin{tabular}[|c|]{@{}c@{}} 901.6 \\ 30.1 \\ 839.0 \end{tabular}  
& \begin{tabular}[|c|]{@{}c@{}} 1415.5 \\ 300.3 \\ 974.4 \end{tabular}  
\\ \midrule
\textbf{b6}&\begin{tabular}[|c|]{@{}c@{}}$T_{\text{E2E}}$\\ $T_{\text{LoC}}$ \\  $T_{\text{LoH}}$\end{tabular}
& \begin{tabular}[|c|]{@{}c@{}} 3.139 \\ 0.249 \\ 1.330 \end{tabular}  
& \begin{tabular}[|c|]{@{}c@{}} 1.201 \\ 0.258 \\ 0.453 \end{tabular}  
& \begin{tabular}[|c|]{@{}c@{}} 3.24 \\ 0.58 \\ 1.38 \end{tabular}  
& \begin{tabular}[|c|]{@{}c@{}} 17.69 \\ 2.69 \\ 8.99 \end{tabular}  
& \begin{tabular}[|c|]{@{}c@{}} 219.2  \\ 50.0 \\ 107.6 \end{tabular}  
& \begin{tabular}[|c|]{@{}c@{}} 123.1 \\ 18.7 \\ 71.9 \end{tabular}  
& \begin{tabular}[|c|]{@{}c@{}} 680.9 \\ 270.9 \\ 269.2 \end{tabular}  
\\ \midrule
\textbf{b7}&\begin{tabular}[|c|]{@{}c@{}}$T_{\text{E2E}}$\\$T_{\text{LoC}}$ \\ $T_{\text{LoH}}$ \end{tabular}
& \begin{tabular}[|c|]{@{}c@{}} 2.252 \\ 0.223 \\ 0.469 \end{tabular}  
& \begin{tabular}[|c|]{@{}c@{}} 0.826 \\ 0.235 \\ 0.101 \end{tabular}  
& \begin{tabular}[|c|]{@{}c@{}} 2.285 \\ 0.594 \\ 0.411 \end{tabular}  
& \begin{tabular}[|c|]{@{}c@{}} 11.32 \\ 2.63 \\ 2.68\end{tabular}  
& \begin{tabular}[|c|]{@{}c@{}} 368.8 \\ 53.8 \\ 253.4  \end{tabular}  
& \begin{tabular}[|c|]{@{}c@{}} 72.1 \\ 17.5 \\ 22.1 \end{tabular}  
& \begin{tabular}[|c|]{@{}c@{}} 601.8 \\ 261.4 \\ 199.6 \end{tabular}  
\\ \midrule
\textbf{b8}&\begin{tabular}[|c|]{@{}c@{}} $T_{\text{E2E}}$ \\$T_{\text{LoC}}$  \\  $T_{\text{LoH}}$ \end{tabular}
& \begin{tabular}[|c|]{@{}c@{}} 7.98 \\ 0.23 \\ 6.19 \end{tabular}  
& \begin{tabular}[|c|]{@{}c@{}} 3.25 \\ 0.24 \\ 2.52 \end{tabular}  
& \begin{tabular}[|c|]{@{}c@{}} 13.79 \\ 0.61 \\ 11.90 \end{tabular}  
& \begin{tabular}[|c|]{@{}c@{}} 67.65 \\ 2.74 \\ 58.90 \end{tabular}  
& \begin{tabular}[|c|]{@{}c@{}} 537.8 \\ 52.2 \\ 424.0 \end{tabular}  
& \begin{tabular}[|c|]{@{}c@{}} 548.2 \\ 28.7 \\ 487.0 \end{tabular}  
& \begin{tabular}[|c|]{@{}c@{}} 1749.3 \\ 283.5 \\ 1325.0 \end{tabular}  
\\ \bottomrule
\end{tabular}
\end{adjustbox}
\label{tab:measured-latency-GraphAGILE}
\end{table}
\end{scriptsize}

\begin{scriptsize}
\begin{table}[ht]
\centering
\caption{The size (MB)  of the generated binary files [Row 1-8], and the size (MB)  of input graphs [Row 9]}
\begin{adjustbox}{max width=0.45\textwidth}
\begin{tabular}{ccccccccc}
\toprule
 & \multicolumn{7}{c}{\textbf{Dataset}} \\ \cmidrule{2-8}
& CI&CO&PU&FL&RE&YE&AP\\ \midrule
\textbf{b1}
& 0.136  
& 0.053  
& 0.193  
& 0.194 
& 0.228  
& 0.161 
& 0.246 
\\ \midrule
\textbf{b2}
& 0.141 
& 0.057  
& 0.234  
& 0.270  
& 0.234  
& 0.218  
& 0.369  
\\ \midrule
\textbf{b3}
& 0.210 
& 0.084  
& 0.340  
& 0.393 
& 0.340 
& 0.310 
& 0.518 
\\ \midrule
\textbf{b4}
& 0.217 
& 0.093 
& 0.421 
& 0.421  
& 0.427 
& 0.423 
& 0.764 
\\ \midrule
\textbf{b5}
&  0.297 
&  0.131 
&  0.632 
&  0.633 
&  0.703 
&  0.661 
&  1.231 
\\ \midrule
\textbf{b6}
&  0.145 
&  0.060 
&  0.263 
&  0.299 
&  0.264 
&  0.258 
&  0.457 
\\ \midrule
\textbf{b7}
&  0.204 
&  0.079 
&  0.281 
&  0.281 
&  0.334 
&  0.230 
&  0.342 
\\ \midrule
\textbf{b8}
&  0.101
&  0.059 
&  0.422 
&  0.422 
&  0.439 
&  0.528 
&  1.098 
\\ \midrule
\textbf{Input graph}
&  47 
&  12.6 
&  38 
&  181 
&  1863 
&  900 
&  4223 
\\ \bottomrule
\end{tabular}
\end{adjustbox}
\label{tab:size-of-binary}
\end{table}
\end{scriptsize}

\noindent \textbf{Execution time}: Table \ref{tab:measured-latency-GraphAGILE} shows the measured latency of GraphAGILE. 
We observe that the software compilation time ranges from $2$ ms to  $300$ ms, which is proportional to the size of the input graph. The reason is that data partitioning is the most time-consuming operation with complexity $\mathcal{O}(|\mathcal{V}| + |\mathcal{E}|)$. The design automation frameworks (e.g., DeepBurning-GL \cite{liang2020deepburning}) undergo hours of overhead to perform hardware synthesis and Place\&Route. Thus, the proposed software compiler is fast and lightweight. 

\vspace{0.1cm}
\noindent \textbf{Size of binary file}: Table \ref{tab:size-of-binary} shows the size of the generated binary files. Compared with the sizes of input graphs or the inter-layer intermediate results, the size of binary files is negligible. Therefore, loading the binary files from the FPGA external DDR memory to the on-chip scheduler results in a small amount of memory traffic. The size of the binary files is small because the high-level instructions are compact and powerful; For example, a single high-level instruction (128 bits) can define the computation task of a large data partition (up to 16384 vertices).

\subsection{Impact of the Optimizations}
\label{subsec:Impact-of-the-Optimizations}
\begin{figure*}[h]
     \centering
     \includegraphics[width=18cm]{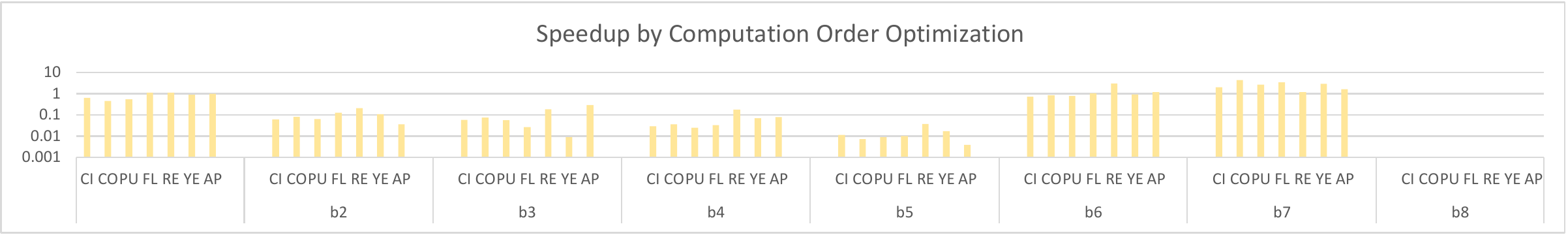}
     \caption{Impact of computation order optimization on the latency of hardware execution (LoH) 
     $T_{\text{LoH}}$} \label{fig:impact-COO}
     \vspace{0.2CM}
     \includegraphics[width=18cm]{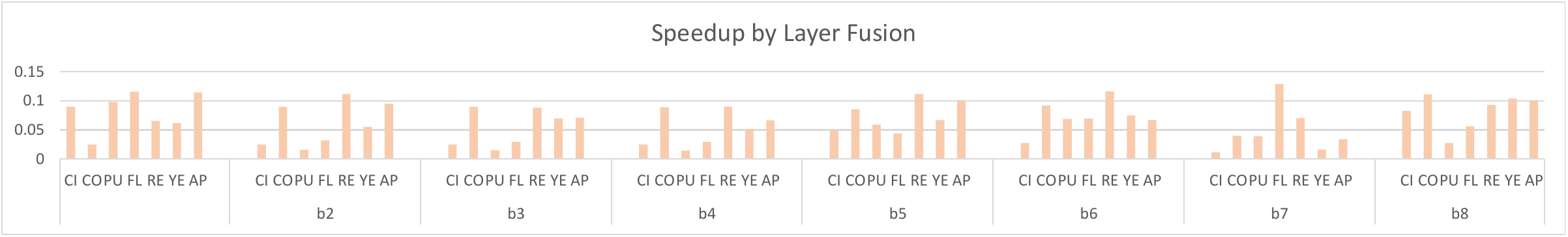}
     \caption{Impact of layer fusion on the latency of hardware execution (LoH) $T_{\text{LoH}}$} \label{fig:impact-LF}
      \vspace{0.2CM}
     \includegraphics[width=18cm]{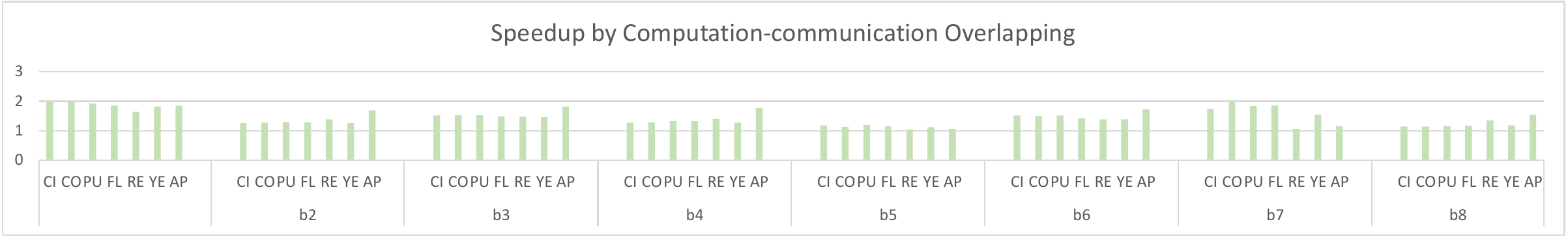}
     \caption{Impact of computation and communication overlapping on the latency of hardware execution (LoH) $T_{\text{LoH}}$}
     \label{fig:impact-CCO}
\end{figure*}

To show the effectiveness of the proposed optimizations, we compare $T_{\text{LoH}}$ of using the compiler optimizations and  $T_{\text{LoH}}$ without compiler optimizations. Figure \ref{fig:impact-COO}, Figure \ref{fig:impact-LF}, and Figure \ref{fig:impact-CCO} show the impact of (1) computation order optimization, (2) layer fusion, and (3) overlapping the computation and data communication (in task scheduling), respectively.

\vspace{0.1cm}
\noindent \textbf{Computation order optimization}: Computation order optimization leads to $82\%$, $9.6\%$, $9.9\%$, $6.3\%$, $1.3\%$, $121\%$, $260\%$, $0\%$ average speedup on \textbf{b1-b8}, respectively. The computation order optimization can reduce both the computation complexity and external memory traffic of the involved Aggregate layers. The computation order optimization has no effect on model \textbf{b8}, because model \textbf{b8} uses a preprocessing MLP layer to transform the feature vectors to a uniform length, which eliminates the opportunities for computation order optimization. Note that the Computation order optimization itself has a small overhead ($\approx 0.5\mu s$ average latency) during the software compilation.

\vspace{0.1cm}
\noindent \textbf{Layer fusion}: Layer fusion leads to $8.1\%$, $6.0\%$, $5.5\%$, $5.2\%$, $7.3\%$, $7.4\%$, $4.7\%$, $8.2\%$ average speedup on \textbf{b1-b8}, respectively. The performance improvement is because the individual Activation layers and BatchNorm layers are eliminated (See Section \ref{subsec:layer-fusion}). Thus, extra memory traffic of the Activation and BatchNorm layers is eliminated to reduce the latency of hardware execution. Note that layer fusion has complexity  $\mathcal{O}(L)$ and incurs small overhead  ($\approx 0.66\mu s$ average latency) during the software compilation.

\vspace{0.1cm}
\noindent \textbf{Overlapping computation and communication}: Overlapping the computation and communication leads to $186\%$, $134\%$, $153\%$, $137\%$, $112\%$, $148\%$, $158\%$, $123\%$ average speedup on \textbf{b1-b8}, respectively. It demonstrates the effectiveness of proposed double/triple buffering techniques and the effectiveness of the software compilation optimizations.

\begin{figure*}[h]
     \centering
     \includegraphics[width=18cm]{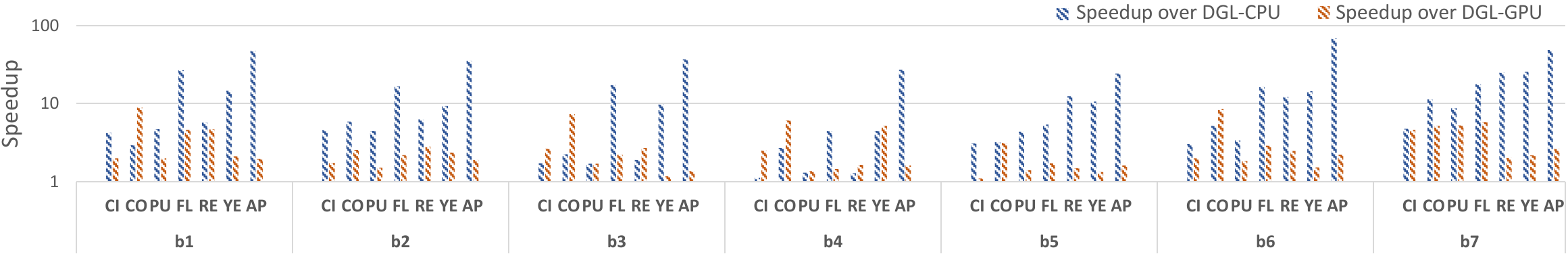}
     \caption{Comparison of end-to-end latency $T_{\text{E2E}}$ with DGL}
     \label{fig:end-to-end-cmp-DGL}
\end{figure*}

\begin{figure*}[h]
     \centering
     \includegraphics[width=18cm]{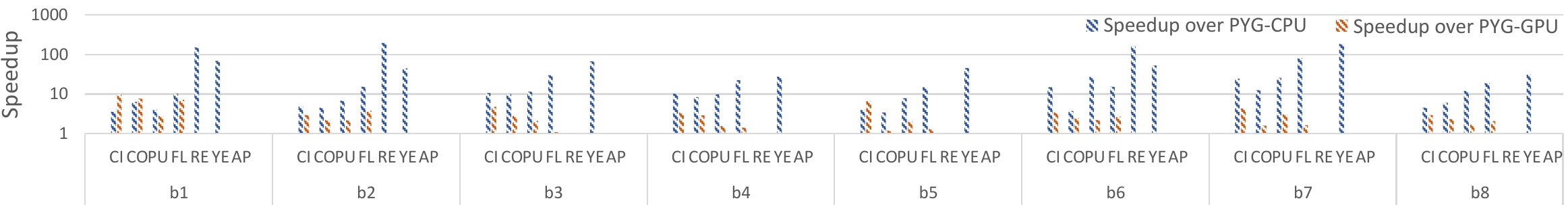}
     \caption{The comparison of end-to-end latency $T_{\text{E2E}}$ with PyG. Note that PyG-CPU cannot execute AP due to out of memory. PyG-GPU cannot execute RE, YE, and AP due to out of memory. Therefore, these results are not shown in the Figure.}
     \label{fig:end-to-end-cmp-PYG}
\end{figure*}


\subsection{Cross Platform Comparison}
\label{subsec:cross-platform-comparsion}

We compare $T_{\text{E2E}}$ on three baseline platforms: (1) CPU-only platform, (2) CPU (Ryzen 3990x) + GPU, (3)  CPU (Ryzen 3990x) + GraphAGILE. On CPU-only platform, we execute CPU version of Pytorch Geometric (PyG) and Deep Graph Library (DGL), with Intel MKL as the backend. On CPU-GPU platform, we execute GPU version of PyG and DGL, with CUDA 11.3 as the backend. 
The E2E latency of CPU-only and CPU-GPU platforms include the preprocessing overhead of runtime systems (e.g., GPU kernel launch). 
Figures \ref{fig:end-to-end-cmp-DGL} and \ref{fig:end-to-end-cmp-PYG} show the comparison. 
Compared with PyG-CPU, GraphAGILE achieves {$10.3\times-~47.1\times$}  speedup on \textbf{b1-b8}. Compared with PyG-GPU, GraphAGILE achieves {$1.27\times-~3.8\times$}  speedup on \textbf{b1-b8}. Compared with DGL-CPU, GraphAGILE achieves {$9.1\times-~20.1\times$} speedup on \textbf{b1-b7}.  Compared with DGL-GPU, GraphAGILE achieves {$1.7\times-~3.9\times$} speedup on \textbf{b1-b7}. 

The speedup over CPU-only and CPU-GPU platforms is due to: (1) The kernels in GNN (e.g., SpDMM, SDDMM) have irregular computation\&memory access patterns and low data reuse. GraphAGILE hardware architecture optimizes the data path and memory organization for various GNN computation kernels. The processors in CPU or GPU have limited cache sizes (e.g., 32KB L1 cache and 512KB L2 cache).
The data exchange (due to low data reuse) among L1, L2, and L3 caches becomes the performance bottleneck and results in reduced sustained performance. On CPU platforms, loading data from the L3 cache incurs latency of $32$ns, and loading data from L2 cache incurs latency of $5-12$ns. Compared with the CPU-only/CPU-GPU, the ACK in GraphAGILE can access data in one clock cycle from the on-chip edge/weight/feature buffers. Therefore, although the baseline CPU-only and CPU-GPU platforms have higher ($6\times$) peak performance than the state-of-the-art FPGAs, GraphAGILE still outperforms the baselines. (2) The compiler of GraphAGILE automatically performs various optimizations to minimize execution time. While the computation order optimization and layer fusion can potentially be applied to  CPU-only and CPU-GPU platforms, other compiler optimizations (such as data partitioning for partition-centric execution schemes, task scheduling for dynamic load balancing) are specific to the proposed overlay architecture. For example, data partitioning relies on an effective and customized memory organization. The hardware architecture and the compiler of GraphAGILE perform synergistically to achieve lower latency.

\subsection{Comparison with the State-of-The-Art Accelerators}
\label{subsec:Comparison-State-of-The-Art}

\begin{table*}[ht]
\centering
\caption{Advantages of GraphAGILE over the state-of-the-art work}
\label{tab:advantages}
\begin{threeparttable}[t]
\begin{tabular}{lcccccc}
\toprule
 & GAT & NHC $\star$  & Preprocesssing & UFH $\ddagger$ & GEMM & SDDMM \\
 \midrule
{HyGCN \cite{yan2020hygcn}} & No \Frowny{} & No \Smiley{}& graph partitioning, sparsity elimination  &  No \Frowny{} & YES \Smiley{}  & NO  \Frowny{}\\
{AWB-GCN \cite{geng2020awb}} & No \Frowny{} & No \Smiley{} & graph partitioning, data layout transformation  & YES \Smiley{} &  NO \Frowny{} & NO  \Frowny{}\\
{DeepBurning-GL \cite{liang2020deepburning}} & No \Frowny{} & Yes (6-8 hours)  \Frowny{} &  (Unknown)  & NO \Frowny{} & YES \Smiley{}  & NO \Frowny{}\\
{BoostGCN \cite{zhang2021boostgcn}} & No \Frowny{} & Yes (6-8 hours)  \Frowny{} & graph partitioning  &  NO \Frowny{} &  YES \Smiley{}  &  NO \Frowny{} \\
GraphAGILE & YES \Smiley{}  &  No \Smiley{} & software compilation  &  YES \Smiley{} &  YES \Smiley{}  & YES  \Smiley{} \\
 \bottomrule
\end{tabular}
\begin{tablenotes}
\item[] $\star$ NHC: if the design needs to regenerate hardware if the GNN model or input graph is changed.
\item[] $\ddagger$ UFH: if the design uses the unified hardware module to execute various computation kernels.
 \end{tablenotes}
\end{threeparttable}
\end{table*}

We compare with state-of-the-art accelerators: HyGCN \cite{yan2020hygcn}, AWB-GCN \cite{geng2020awb}, DeepBuring-GL \cite{liang2020deepburning} and BoostGCN \cite{zhang2021boostgcn}.

\vspace{0.1cm}
\noindent \textbf{Advantages of GraphAGILE}: Table \ref{tab:advantages} summarizes the performance comparison. HyGCN \cite{yan2020hygcn} and AWB-GCN \cite{geng2020awb} use fixed hardware designs that only support limited GNN models. For example, they cannot execute GAT due to the lack of support for SDDMM. Moreover, they use additional data-dependent optimizations, such as sparsity elimination (HyGCN). These optimizations can reduce the latency of hardware execution $T_{\text{LoH}}$ at the cost of increased end-to-end latency due to the expensive preprocessing. 
Design automation frameworks such as 
DeepBurning-GL \cite{liang2020deepburning} and BoostGCN \cite{zhang2021boostgcn} 
need to pay hours of overhead to regenerate FPGA bitstream for every pair of GNN models and input graph.  
Therefore, they have very large end-to-end latency.  
HyGCN, DeepBurning-GL, and BoostGCN are hybrid architectures that 
initialize different hardware modules for various computation kernels. 
However, hybrid architectures suffer from load imbalance and thus, hardware under-utilization. AWB-GCN uses the same set of processing elements to execute SpDMM under various data sparsity. It is not efficient for GEMM and does not support SDDMM. For dense input graphs (e.g., AmazonProducts) or GNN models with the PReLU or SWISH activation functions, GEMM is essential to be supported.

\begin{table}[]
\centering
\caption{Comparison of $T_{\text{LoH}}$}
\label{tab:com-pure-inference-latency}
\begin{adjustbox}{max width=0.45\textwidth}
\begin{tabular}{ccccc}
\toprule
\textbf{Model}         & 
\textbf{Dataset} & \textbf{Approach} &  $T_{\text{LoH}}$ (ms) & \textbf{Speedup}\\
\midrule
\multirow{10}{*}{\textbf{b2}} &FL&
\begin{tabular}[|c|]{@{}c@{}} BoostGCN \cite{zhang2021boostgcn} \\ GraphAGILE  \end{tabular}        
&  \begin{tabular}[|c|]{@{}c@{}} 20.1 \\ 11.5 \end{tabular}   
&  \begin{tabular}[|c|]{@{}c@{}} $1.75\times$  \\ $1\times$ \end{tabular}  
\\ \cmidrule{2-5}
&RE&\begin{tabular}[|c|]{@{}c@{}} BoostGCN \cite{zhang2021boostgcn} \\ HyGCN \cite{yan2020hygcn}\\ AWB-GCN \cite{geng2020awb} \\ GraphAGILE  \end{tabular} &
 \begin{tabular}[|c|]{@{}c@{}} 98.1 \\ 289 \\ 49.7 \\  97.2 \end{tabular}  
 &  \begin{tabular}[|c|]{@{}c@{}} $1.01\times$ \\ $2.97\times$ \\ $0.51\times$  \\ $1\times$ \end{tabular} 
\\ \cmidrule{2-5}
&YE&\begin{tabular}[|c|]{@{}c@{}} BoostGCN \cite{zhang2021boostgcn} \\ GraphAGILE  \end{tabular}& 
\begin{tabular}[|c|]{@{}c@{}} 193 \\ 104.3  \end{tabular} 
&  \begin{tabular}[|c|]{@{}c@{}} $1.85\times$  \\ $1\times$ \end{tabular}  
\\ \cmidrule{2-5}
&AP&\begin{tabular}[|c|]{@{}c@{}} BoostGCN \cite{zhang2021boostgcn} \\ GraphAGILE  \end{tabular} & 
\begin{tabular}[|c|]{@{}c@{}} 793.5 \\ 315.9  \end{tabular}
&  \begin{tabular}[|c|]{@{}c@{}} $2.51\times$  \\ $1\times$ \end{tabular} 
\\ \bottomrule
\end{tabular}
\end{adjustbox}
\end{table}

\vspace{0.1cm}
\noindent \textbf{Comparison of latency of hardware execution $T_{\text{LoH}}$}: Since no previous work measure the end-to-end latency, their overhead of graph preprocessing (Table \ref{tab:advantages}) are unknown. Therefore, we are only able to compare the latency of hardware execution $T_{\text{LoH}}$, as shown in Table \ref{tab:com-pure-inference-latency}.  Table \ref{tab:resource-utilizations}  shows the detailed resource utilization of various FPGA accelerators. Compared with BoostGCN, GraphAGILE achieves $1.01\times-~2.51\times$ speedup on FL, RE, YE, and AP under comparable peak performance and memory bandwidth. Compared with HyGCN, GraphAGILE achieves $2.97\times$ speedup on RE. GraphAGILE achieves higher performance because BoostGCN and HyGCN are hybrid accelerators that suffer from load imbalance.   AWB-GCN is $1.96\times$ faster than GraphAGILE on RE because (1) the platform of AWB-GCN has $2.2\times$ peak performance than GraphAGILE, and (2) AWB-GCN exploits the sparsity of vertex features to reduce the total computation complexity. However, the sparsity exploitation in AWB-GCN requires a runtime system to obtain the sparsity of the intermediate results and dynamically perform data format transformation and kernel remapping. Therefore, the runtime optimizations of AWB-GCN are orthogonal to our static compiler optimizations. For an overlay accelerator, it is challenging to exploit the data sparsity because both data format and high-level instructions need to be generated/changed dynamically at runtime. We leave the dynamic data sparsity optimizations in the runtime system as future work.

\section{Discussion}
{In real-world applications, the input graphs can be very large, consisting of billions of vertices and edges. For example, the ogbn-papers100M \cite{hu2020open} dataset requires more than 100 GB DDR memory to store the full input graph, which is beyond the capacity of the DDR memory on the state-of-the-art FPGA boards (e.g., Xilinx Alveo U250 board has 64 GB on-board DDR memory). GraphAGILE can be easily extended to perform GNN inference on large-scale input graphs. To achieve this, the following are required: (1) coarse-grained data partitioning by the compiler, and (2) a runtime system to perform task scheduling and inter-data-partition communication on the host processor. During compilation, the compiler first partitions the input graph into super data partitions, each fitting in half of the total FPGA on-board DDR memory. Using half of the total FPGA on-board DDR memory, we can overlap the computation and CPU-FPGA data communication via double-buffering. Then, following Section \ref{sec:compiler}, the compiler performs fine-grained data partitioning, kernel mapping, and task scheduling for each super data partition. The compiler will generate a binary file for each super data partition. At runtime, the runtime system on the host processor schedules the execution of super data partitions onto the FPGA accelerator. The runtime system also performs inter-data-partition communication by sending the data from other super data partitions (in the host memory) to the super data partition currently on the FPGA accelerator. The computation on the accelerator and the CPU-FPGA data communication can be overlapped to improve the overall performance. We leave the support for the large-scale input graphs as future work.
}

\section{Conclusion and Future Work}
In this work, we proposed a domain-specific overlay accelerator for low-latency GNN inference. The proposed accelerator consists of a novel hardware architecture with an instruction set, and a software compiler with various optimizations for latency reduction. The experimental results showed that compared with the state-of-the-art implementations on CPU-only and CPU-GPU platforms,  we achieved $47.1\times$ and $3.9\times$ speedup in end-to-end latency. Compared with state-of-the-art GNN accelerators, we achieved up to $2.9\times$ speedup in terms of hardware execution latency. 
 GraphAGILE has supported widely-used GNN models, including the numerous GNN models in GraphGYM.  

 \vspace{0.1cm}
 \noindent \textbf{Future work}: (1) In the future, we intend to extend GraphAGILE to support various GNN minibatch training algorithms and develop a design automation algorithm to quickly generate the overlay accelerator for a given FPGA platform.
 { (2) We also plan to build a runtime system that performs dynamic sparsity exploitation to reduce the hardware execution latency. The runtime system will perform just-in-time (JIT) compilation, which can dynamically map a computation task (e.g., a layer block in Algorithm \ref{algo:PCFA}) to a computation kernel (e.g., GEMM or SpDMM) based on the data sparsity.}


\section*{Acknowledgment}
This work is supported by the National Science Foundation (NSF) under grants CCF-1919289 and OAC-2209563. Equipment and support by AMD Xilinx are greatly appreciated.

\bibliographystyle{IEEEtran}
\bibliography{citation}

\begin{IEEEbiography}[{\includegraphics[width=1in,height=1.25in,clip,keepaspectratio]{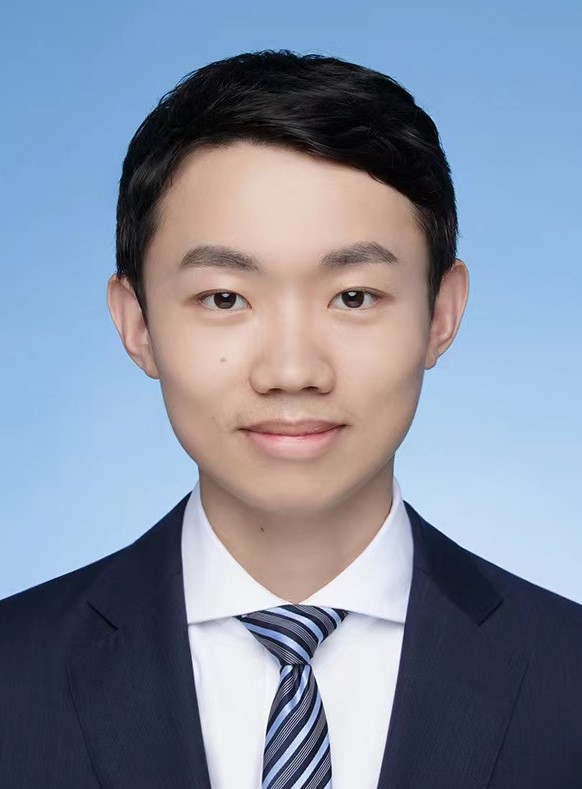}}]{Bingyi Zhang}
received the BS degree in microelectronics from Fudan University in 2017, and the MS degree in Integrated Circuit Engineering from Fudan University. He is pursuing the Ph.D. degree in Computer Engineering at the University of Southern California (USC). His research interests include parallel computing, digital signal processing, digital circuit design. His current work is focused on accelerating graph-based machine learning on FPGA. 
\end{IEEEbiography}
\vskip -3\baselineskip plus -1fil
\begin{IEEEbiography}[{\includegraphics[width=1in,height=1.25in,clip,keepaspectratio]{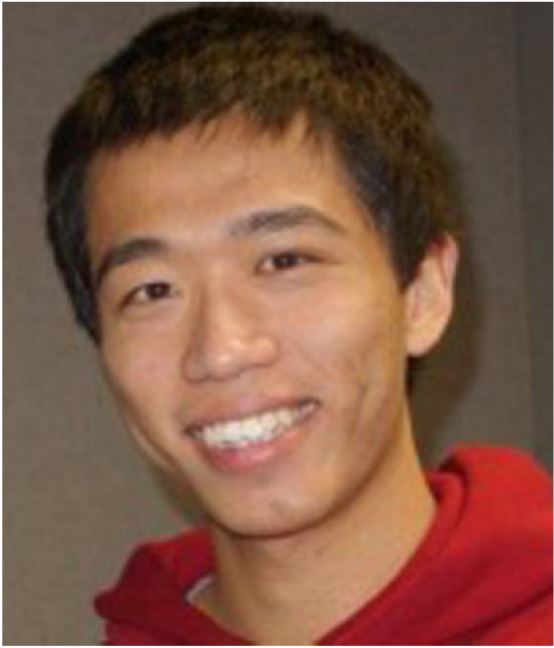}}]{Hanqing Zeng} received the B. Eng degree in electronic engineering from the University of Hong Kong in 2016, and the Ph.D. degree in Computer Engineering from University of Southern California in 2022. He is currently  a research scientist at Meta AI. His research interests include large scale graph representation learning, parallel and distributed computing, and algorithm-architecture co-optimization for deep learning applications.
\end{IEEEbiography}
\vskip -3\baselineskip plus -1fil
\begin{IEEEbiography}[{\includegraphics[width=1in,height=1.25in,clip,keepaspectratio]{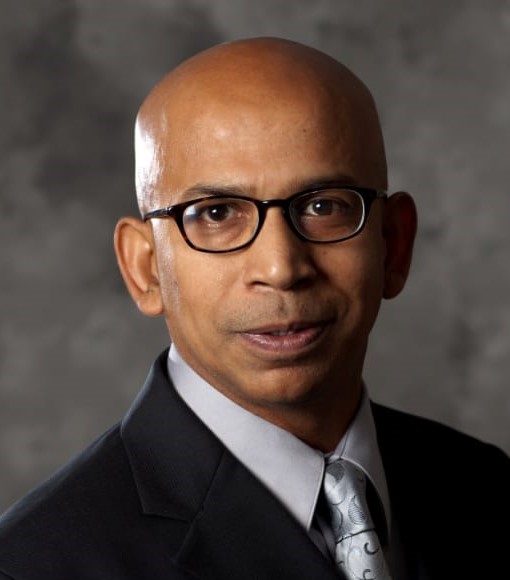}}]{Viktor K. Prasanna}
 received the BS degree in electronics engineering from Bangalore University, the MS
degree from the School of Automation, Indian Institute
of Science, and the Ph.D. degree in computer science from Pennsylvania State University. He is Charles
Lee Powell chair in engineering in the Ming Hsieh
Department of Electrical and Computer Engineering and professor
of computer science at the University of Southern California (USC). His research interests include
high performance computing, parallel and distributed
systems, reconfigurable computing, and embedded systems. He is the executive director of the USC-Infosys Center for Advanced
Software Technologies (CAST) and was an associate director of the USC Chevron
Center of Excellence for Research and Academic Training on Interactive Smart
Oilfield Technologies (Cisoft). He also serves as the director of the Center
for Energy Informatics, USC. He served as the editor-in-chief of the IEEE
Transactions on Computers during 2003–06. Currently, he is the editor-in-chief
of the Journal of Parallel and Distributed Computing. He was the founding
chair of the IEEE Computer Society Technical Committee on Parallel Processing.
He is the steering chair of the IEEE International Parallel and Distributed
Processing Symposium (IPDPS) and is the steering chair of the IEEE International
Conference on High Performance Computing (HiPC). He received the 2009
Outstanding Engineering Alumnus Award from the Pennsylvania State University. He received the W. Wallace McDowell Award from the IEEE Computer
Society, in 2015 for his contributions to reconfigurable computing. His work
on regular expression matching received one of the most significant papers
in FCCM during its first 20 years award in 2013. He is a fellow of the IEEE, the ACM, and the American Association for Advancement of Science
(AAAS). He was recently elected as the member of Academia Europaea.
\end{IEEEbiography}




\clearpage

  
\end{document}